\documentclass[useAMS,usenatbib]{mn2e} 
\usepackage{graphicx}
\usepackage{color}

\title[X-ray properties of Clusters of Galaxies and Bright Galaxies]{Multi-wavelength investigations of co-evolution of bright cluster galaxies and their host clusters}
\author[Hashimoto et al.]{Yasuhiro Hashimoto$^{1}$\thanks{Email:
hashimot@ntnu.edu.tw} J. Patrick Henry$^{2,3}$
 and  Hans Boehringer$^{3}$ \\
$^{1}$Department of Earth Sciences, National Taiwan Normal University
     No.88, Sec. 4, Tingzhou Rd., Wenshan District, Taipei 11677, Taiwan R.O.C \\
$^{2}$Institute for Astronomy, University of Hawaii, 2680 Woodlawn Drive, Honolulu, Hawaii 96822, USA\\
$^{3}$Max-Planck-Institut f\"ur extraterrestrische Physik,
              Giessenbachstrasse
              D-85748 Garching, Germany\\
}

\begin{document}

\date{Accepted 2014 February 14}

\pagerange{\pageref{firstpage}--\pageref{lastpage}} \pubyear{2002}

\maketitle

\label{firstpage}

\begin{abstract}

We report a systematic multi-wavelength investigation of
environments of the brightest cluster galaxies (BCGs),
using the X-ray data from the Chandra archive,
and optical images taken with 34' $\times$ 27' field-of-view
Subaru Suprime-Cam.
Our goal is to help understand
the relationship between the BCGs and their host clusters,
and between the BCGs and other galaxies,
to eventually address a question of
the formation and co-evolution of BCGs and the clusters.

Our results include:
1) Morphological variety of BCGs, or the second or the third brightest galaxy (BCG2, BCG3), is comparable to that of other bright red sequence galaxies,
 suggesting that
 we have a continuous variation of morphology
 between BCGs, BCG2, and BCG3,
 rather than  a sharp separation
 between the BCG and the rest of the bright galaxies.
2) The offset of the BCG position relative to the cluster centre is correlated
to the degree of concentration of cluster X-ray morphology
(Spearman $\rho$ = -0.79),
consistent with an interpretation that BCGs tend to be off-centered inside dynamically unsettled clusters.
3) Morphologically disturbed clusters tend to harbour the brighter BCGs,
implying that the ``early collapse''
may not be the only major mechanism to control the BCG formation and evolution.

\end{abstract}


\section{Introduction}

 It has been known for several decades  that a significant fraction
  of galaxy clusters 
show disturbed cluster morphologies,
indicative of possible recent mergers 
\citep[e.g.][]{geller1982substructure,dressler1988evidence}.
In addition to the obvious interest in the merger phenomena themselves, 
the important connection between the morphologies of galaxy clusters and
  the properties of member galaxies has also received
 much attention
 \citep[e.g.][]{butcher1978egc,caldwell1993sfe,metevier2000boe,gerke2007deep2}.
   This connection has generally been formulated in terms of the frequency
   of `structure' in clusters and
from qualitative measures of the properties of the  member galaxies.

  Methods to quantify cluster structures at optical wavelengths
     have mostly used both the
     distribution of cluster galaxies, and lensing.
     The distribution studies analyze, either visually or objectively, the substructure
     in 1D, 2D or sometimes 3D, depending on the level of information available
    \citep[e.g.][]{geller1982substructure,dressler1988evidence,rhee1992eac,bird1994substructure,kriessler1997substructure}.
     \citet{bautz1970classification}
   classified cluster morphology
    based on visual inspection of
optical images, by the degree to which the brightest cluster
member stands out against the cluster background.
    The RS system 
   \citep[][]{rood1971tuning,struble1984morphological,struble1987compilation}
is based on the projected distribution of the
brightest galaxies in the cluster.
The RS system is composed of six major
classes: cD, B, C, L, F, and I-type clusters.
These six classes have been interpreted as corresponding to a sequence of cluster evolution 
\citep[][]{forman1982x,struble1984morphological}.
  \citet{butcher1984egc}
  characterized the
optical morphology of cluster by the degree of central concentration
 of galaxy distribution.
Their concentration  is defined by log(R$_{60}$/R$_{20}$) where R$_{60}$
or R$_{20}$
is the radius of the circle containing 60\% or 20\% of the cluster
projected galaxy distribution, respectively.

An alternative method comes from X-ray wavelengths, because
     cluster mergers compress and heat the intracluster gas,
     and this can be measured as distortions of the spatial
     distribution of X-ray surface brightness and temperature.
     \citet{jones1999einstein}
     visually examined 208
      clusters observed with {\it Einstein} X-ray satellite and
      separated these clusters into six morphological classes.
  Meanwhile,
   using the $Einstein$ images,
   \citet{mohr1995cosmological}
   measured emission-weighted centroid variation,
   axial ratio, orientation, and radial falloff for a sample of 65 clusters,
    while several other studies  used
     ellipticity 
      \citep[e.g.][]{kolokotronis2001searching,melott2001recent,plionis2002recent}.
      \nocite{buote1995reliability,buote1996quantifyin}
      Buote \& Tsai (1995, 1996) used a
      power ratio method for 59 low redshift clusters observed with $ROSAT$,
       and
      \citet{jeltema2005evolution}
      have extended the method
      to 40 clusters at z=0.15-0.9 using $Chandra$ data.
      \citet{2001AA...378..408S}
      conducted a study of 470 clusters
      from the $ROSAT$-ESO Flux-Limited X-ray (REFLEX) cluster survey
      \citep[][]{2001AA...369..826B}
      using sophisticated statistics, such as Fourier elongation test,
      Lee test, and $\beta$ test.
      \citet{hashimoto2007rqm} 
      studied  X-ray cluster morphology
using a sample of 101 clusters of galaxies
at redshift z$\sim$0.05-1 taken from the Chandra archive.
There, X-ray morphology is quantitatively characterized by
a series of objectively measured simple statistics,
such as concentration, asymmetry, elongation, and off-centerness  of
the X-ray surface brightness distribution.
These measures
are designed to be  robust against variations of image quality
caused by various exposure times and various cluster redshifts.

      Quantifying cluster structures by investigating
      the distribution of cluster galaxies in optical
      wavelengths requires
      a large number of galaxies,
     and is more susceptible to contamination
     from foreground and background objects.
     Lensing study in optical wavelengths is
     also sensitive to this contamination,
     and does not have good spatial resolution except for the
     central region of a cluster.
      The X-ray method is superior against
      fore/background because
      X-ray emissivity is proportional to the square
   of the electron density, and therefore less affected
   by the superposed structures than optical data.
   Meanwhile, the advantage of using optical data is the size
   of the available cluster catalogs, which can be much larger than
   those originating from X-ray data.
   Optical and X-ray characteristics of cluster structures are 
   complementary, and 
   by the systematic comparison between X-ray and optical methods,
   one may calibrate and evaluate different sensitivity and  bias between the optical and X-ray characteristics of clusters 
(e.g. Hashimoto et al. 2007b).

There are numerous studies  investigating the relationship between the 
galaxy properties and
their host clusters (as well as smaller scale environments, such as
the local density). 
The galaxy properties investigated range from the colour and 
morphology to the star formation properties of galaxies
\citep[e.g.][]{butcher1978egc,dressler1980galaxy,hashimoto1998ies,hashimoto1999concentration,goto2003morphology}.
Unfortunately, most of these studies only uses the optical band to
characterize the clusters. 
Only a handful of studies uses multi-band information from both
optical and X-ray data
\citep[e.g.][]{edge1991exosat,metevier2000boe,hashimoto2008aga}.
These studies are important, however, a more systematic investigation of cluster
galaxies using multi-wavelength data in a coherent manner is much
needed. 

Among all cluster galaxies, 
the brightest cluster galaxies (BCGs) are a unique class of objects.
Despite their apparent morphological resemblance to elliptical galaxies,
BCGs tend to have lower surface brightness \citep[e.g.][]{vonderlinden2007},
while their spatial extend is much larger (effective radius $\sim$ 30 kpc) 
than ordinary elliptical galaxies \citep[e.g.][]{schneideretal1983,schombert1986,gonzalezetal2005}.
Meanwhile, BCGs tend to have smaller velocity dispersions and smaller colour gradients in their radial profiles 
\citep[][]{bernardi2011evidence}.
BCGs also tend to lie close to the center of cluster in both 2-d and velocity 
space 
\citep[e.g.][]{quintana1982determination,oegerle2001dynamics},
implying that
they are often located at the minimum in the cluster potential well,  
thus their formation history may have been  dominated by different physical 
processes compared to the other galaxies in the clusters.

According to the cold dark matter model, 
BCGs form hierarchically by the merging of smaller galaxies, and
the formation history of the BCG is closely linked to that of the
host cluster \citep[e.g.][]{de2007hierarchical}. 
Indeed, observationally, it has been found that BCGs' properties are 
closely related to those of host clusters: a significant alignment 
between the elongations of BCGs and their host clusters
is observed in both the optical \citep[][]{carter1980mcg,struble1990pas,plionis2003gap} and X-ray bands \citep[][]{hashimoto2008aga}.
The correlation between the BCG luminosity and various optical cluster properties are also
found \citep[e.g.][]{oemler1976structure,schombert1987structure,lin2004k}.
The BCG luminosity is also found to be weakly correlated to cluster X-ray temperature or luminosity 
\citep[e.g.][]{schombert1987structure,edge1991exosat,brough2002evolution,katayama2003properties}.

The unique characters of BCGs allows us to use them as
important diagnostics of the dynamical status of the host clusters.
One of the most important example is that
the offset between the BCG and the center of global cluster
potential well is expected to be 
sensitive to the cluster dynamical state 
\citep[e.g.][]{ostriker1975another,merritt1985distribution,katayama2003properties}.
\citet{beergeller1983} investigated the offset of cD galaxy from the peak of the 
galaxy surface density, using a sample of 55 nearby rich clusters of galaxies.
They found that cD galaxies often do not lie at the 
global center of the galaxy surface distribution.
They 
suggested that
these galaxies tend to lie at the bottom of {\em local} potential wells,
rather than global potential wells.
The offset of cD (and D) galaxies with respect to the host clusters
in velocity space is also reported   
\citep[][]{malumuth1992dynamics,zabludoff1993kinematics,bird1994substructure,oegerle2001dynamics}.

These results suggest that BCGs may not always be at the bottom of the 
global potential well of their host clusters. 
However, these previous studies quantifying cluster structures by the galaxy surface
density were unfortunately  sensitive to fore/background contamination.
Even if one uses spectroscopic/photometric redshifts 
to help identify cluster members,
determining the center from the galaxy surface density often lacks
the spatial resolution, because the redshift information (particularly,
those of spectroscopic redshifts) are not provided for all cluster galaxies. Even if 
we could have complete cluster membership information for all galaxies, we still have an intrinsically  
limited spatial resolution because  each galaxy itself can act as a discrete noise, particularly after discarding the BCG from the galaxy
distribution.

X-ray is superior for determining cluster structure, such as the center
of the cluster, because it is  less affected by the superposed structure than optical bands. 
Several researchers 
\citep[e.g.][]{katayama2003properties,sanderson2009locuss}
investigated the offset between
BCG and X-ray centroid for nearby clusters, and confirmed a correlation between 
the BCG offset and presence of cool core, or radio emission.
Unfortunately, most previous studies are predominantly
using the cluster sample that consists of low redshift 
bright clusters, that preferentially contains dynamically settled clusters.
Therefore, the effect of cluster dynamical status on the luminosity (or other
properties) of BCGs were hardly addressed. 
Furthermore,
despite the large number of previous work,
the nature of formation of the BCGs, in particular,
whether or not the BCGs are special, rather than representing
the extreme bright end of the normal galaxy populations, is an open question.
It is unclear whether or not various  `features' discovered
among BCGs can be explained as `continuous' extension of normal galaxies,
or represent discrete signature to separate BCGs from other galaxies.

Unfortunately, in the large majority of previous studies comparing
the properties of clusters and BCGs or generic cluster member galaxies,
the cluster properties were characterized predominantly in optical/NIR bands 
based on the galaxy surface density. 
Even if the X-ray was additionally used to characterize cluster properties,
they are typically just global X-ray luminosity or temperature,
and no detailed X-ray analysis was conducted.
Meanwhile, for a small number of studies where 
the X-ray was the `main' waveband used to characterize  the detailed {\em cluster} properties, complementary optical/NIR datasets, if any, for characterizing BCGs and other galaxies were often shallow, and
 assembled from observations with several telescopes.
Finally,
in these `X-ray main' studies, clusters are typically selected from 
the nearby X-ray bright cluster sample  and/or dynamically settled clusters.
Therefore, the effect of cluster dynamical status on the luminosity (or other
properties) of BCGs could not be well addressed. 

What is needed is to investigate cluster properties
and member galaxies in a coherent manner
across the wavebands,
using homogeneous 
datasets covering a wide range of cluster properties.
The optical datasets should ideally be simultaneously wide and deep, while the X-ray data should be both sensitive and
with good spatial resolution so that one can perform detailed dynamical
analysis of distant clusters,  based on the X-ray cluster morphologies, free from the effect of contaminating point sources.

We are conducting 
a new investigation of 
the relationship  between the
cluster properties and the member galaxy properties,
where the properties of galaxies are characterized 
based on optical
images taken with the large field of view 
Suprime-Cam \citep{miyazaki1998cam} on the Subaru 8m telescope,
while cluster properties
are determined by 
the high spatial resolution X-ray data
taken from the Chandra ACIS archive.
We will conduct a systematic multi-band investigation
to study the relationship between the 
clusters and their member galaxies in a wide range of
cluster properties, such as dynamical status, cluster redshifts, and
in a wide range of galaxies properties, 
from core to outskirts, and 
from bright to faint galaxies.

In this paper, we report
the investigation of relationship between cluster properties and 
the brightest cluster galaxies (BCGs).
Our goal is
to help understand
the formation and evolution of BCGs,
including the relationship between
the BCGs and their host clusters, 
and between the BCGs and other galaxies, 
to eventually address a question of 
whether or not the BCGs are special, rather than representing
the extreme bright end of the normal galaxy populations.

This paper is organized as follows. In Sec. 2 \& 3, we describe our sample
and our measures, and
in Sec. 4, systematics and  deblending are described, and 
Sec. 5 summarizes our results.
Throughout the paper, we use $H_{o}$ = 70 km s$^{-1}$ Mpc$^{-1}$,
$\Omega_{m}$=0.3, and $\Omega_{\Lambda}$=0.7, unless otherwise stated.

\section[]{X-ray data and measures}

Here we briefly summarize our sample, X-ray data preparation,
and X-ray measures.
We have used a sample defined in \citet{hashimoto2007rqm}
where 
almost all clusters are selected from flux-limited X-ray surveys, and X-ray
data are
taken from the Chandra ACIS archive.
A lower limit of z = 0.05 or 0.1 is placed on the redshift to ensure that
a cluster is observed with sufficient field-of-view with ACIS-I or ACIS-S, respectively.
The majority of our sample comes from
the $ROSAT$ Brightest Cluster Sample
\citep[BCS;][]{ebeling1998rbc} and the
Extended $ROSAT$ Brightest Cluster Sample \citep[EBCS;][]{ebeling2000rbc}.
To extend our sample to higher redshifts,
additional high-z clusters are selected from various deep  surveys including:
$ROSAT$ Deep Cluster Survey \citep[RDCS;][]{rosati1998rdc},
$Einstein$ Extended Medium Sensitivity Survey \citep[EMSS;][]{gioia1990eoe,henry1992ems}, and
160 Square Degrees $ROSAT$  Survey \citep{vikhlinin1998cgc}.

The resulting sample contains 120 clusters.
At the final stage of our data processing, to employ our full analysis,
we further applied a selection based on the total counts of cluster emission,
eliminating clusters with very
low signal-to-noise ratio.
Clusters whose center is too close to the edge of the ACIS CCD are also removed.
The resulting final sample contains
101 clusters with redshifts between 0.05 - 1.26 (median z = 0.226).
We reprocessed the level=1 event file retrieved from the archive.
The data were filtered to include only the standard
event grades 0,2,3,4,6 and status 0,
then multiple pointings were merged, if any.
We eliminated time intervals of high background count rate
by performing a 3 $\sigma$ clipping of the
background level.
We corrected the images for exposure variations across the field of view, detector response and telescope vignetting.

We detected point sources using the CIAO routine
celldetect with a signal-to-noise threshold for source detection of three.
 An elliptical background annulus region was defined around each source such that
 its outer major and minor axes were three times the size of the source region.
 We removed point sources, except for those at the center of the cluster which was
 mostly the peak of the surface brightness distribution rather than a real point source.
The images were then smoothed  with Gaussian $\sigma$=5".
We decided to use isophotal contours to characterize
an object region, instead of a conventional
circular aperture, because we did not want to introduce any
bias in the shape of an object.
To define constant metric scale to all clusters,
we adjusted an extracting threshold in such a way that
the square root of the detected object area times a constant was 0.5 Mpc,
i.e. const$\sqrt{area}$ = 0.5 Mpc.
We chose const =1.5, because
the isophotal limit of a detected object was best represented by
this value.

Morphology of individual cluster is objectively
characterized by measures, such as,
ellipticity, asymmetry, and concentration.
The ellipticity is  defined by  the ratio of semi-major and semi-minor
axis.
The asymmetry
is measured by first rotating an image by 180 degrees around the
object center, then subtracting the rotated image from the original unrotated
one. The residual signals above zero are summed and then normalized.
The degree of concentration of the surface brightness profile 
  is defined by the ratio
  between central 30\% and whole 100\% elliptical apertures.
\citep[For further detail of the sample and measures, please see ][]{hashimoto2007rqm}.

\section[]{Optical data and measures}

  The optical broad band images taken with
  Suprime-Cam 
   on the Subaru telescope,
   were retrieved from Subaru-Mitaka-Okayama-Kiso Archive (SMOKA).
  Reduction software developed by \citet{yagi2002lfn}
   was used
  for flat-fielding, instrumental distortion correction, differential
  refraction,
  sky subtraction, and stacking.
The camera covers a 34' $\times$ 27' field of view with a
pixel scale of 0\farcs202.
The photometry is calibrated to Vega system using 
SDSS dr8, 
and  transformation from 
\citet{jordi2006empirical}
were used to obtain the zero points. 
For observations without corresponding SDSS data, 
Landolt standards \citep{landolt1992ups}, if any,
were used. 
Data taken under possible non-photometric conditions (estimated by the derived magnitude zero-point versus exposure time plane) are discarded unless 
we can perform direct calibration using standard stars
in and all over the same field of view.
Accuracy of photometry is approximately 0.1 mag. 
The Galactic extinction is corrected using the extinction map of
\citet{schlegel1998mdi}.
K-correction is calculated based on the polynomial approximation
of \citet{chilingarian2010analytical}, where they compare
their approximation to spectral based k-correction by \citet{roche2009spectral}  
and SED based k-correction by KCORRECT \citep{blanton2007k} for a consistency check.
The k-correction for the bulk of our BCG is less than 0.5, 0.4, and
0.25, respectively for R, I+, and Z band.
Extrapolating the polynomial to z$\sim$ 1.0 may require
a bit of care, although in 'redder' bands, such as R, I, and Z, k-correction
behaves relatively well even at these redshifts.
As a precaution, however, we compare the polynomial to another analytical
approximation for E/SO galaxies in R band by \citet{jorgensen1992ccd}
in a form of k$\sim$2.5(1+z) for consistency.
K-correction for z$\sim$1 is about $\sim$ 0.7.
Based on the residuals  \citet{chilingarian2010analytical},
we roughly estimate our k-correction uncertainty to be
about 0.1, 0.05, and 0.05, respectively in R, I+, and Z band for our early type galaxies at z $<$ 0.5,
and about 0.2, 0.1, and 0.1 for the galaxies at z $\sim$ 1.

  We refine the original astrometry written as WCS keyword in the
  distributed archival data
  using the USNO-A2 catalog with positional uncertainties
  less than  $\sim$ 0.2''.
The data were taken under various seeing conditions,
and we used only
images with less than $\sim$ 1\farcs2 seeing.
The optical data retrieved from SMOKA contains 66 clusters
with redshifts between 0.08 - 1.13.

Objects are detected in the `detection' band for each cluster, which is 
determined by the cluster redshift and data availability.
Total global effective exposure time after the filtering,
the detection wavebands, and other information are summarized in Table 1.
Object detection is performed  using
SExtractor \citep[][]{bertin1996sss}.
We used MAG\_AUTO for total magnitudes.
and 2''  aperture magnitudes for colours, if
applicable. 
Star-galaxy separation is performed on the basis of
CLASS\_STAR versus total magnitude diagram.

We selected BCG as the brightest in the detection band inside a projected radius of 0.5 Mpc from the X-ray center.
BCG is then visually inspected and the assignment is, if it is necessary, 
adjusted, in light of the BCG morphology and distance to the X-ray center.
The morphology of BCGs and other generic galaxies are objectively
characterized in a homogeneous fashion 
using similar measures as the X-ray analysis,
including ellipticity, asymmetry, and concentration.
Note that
the concentration used in the X-ray analysis is optimized  to cluster X-ray
profile, therefore, if it is applied to the galaxy profile,
it is too sensitive to the outer faint structure, and thus to the variation
of detection/analysis threshold.
There are many variants of concentration measures characterizing the
galaxy profile 
\citep[e.g.][]{okamuraetal1984,doietal1993,abrahametal1994,hashimoto1998ies,hashimoto1999concentration,conselice2003,goto2003morphology}, where 
the concentration (or `inverse' concentration) is
defined as the ratio of the light inside a certain inner radius to
the light inside a certain outer radius, or
the ratio between the inner and the outer radii.
We have tested  several concentration measures, and
have decided to use  the radius ratio, rather than the flux ratio, taking Petrosian 50 percent radius  as the inner radius,
and  Petrosian  90 percent as the outer radius.
which are relatively robust against various analysis thresholds for the typical galaxy light profile. 
Petrosian 50 percent or 90 percent radius is the radius at which the ratio of the local surface
brightness in an annulus at r to the mean surface brightenss within r is 0.5
or 0.9, respectively \citep[c.f.][]{yasuda2001galaxy}.
The concentration is then defined by 
the outer radius divided by the inner radius, which is
somewhat similar to the {\em inverse} of the `SDSS inverse concentration'
\citep[e.g.][]{goto2003morphology}.
Apart from the concentration index,
there are several parameters that are measured
only for characterizing the galaxy structures:
a `contrast' parameter is measured
by taking the ratio between
the sum of light belonging to the top 30\% brightest pixels
and the total light belonging to an entire object. 
This contrast parameter, when normalized by the concentration,
is designed to measure `blubbiness' of the light distribution,
similar to `Clumpiness' \citep[][]{conselice2003}.
A series of the power ratios \citep[e.g.][]{buote1996quantifying}
that measure the 
square of the ratio
of higher-order multipole moments of the two-dimensional potential to the monopole moment are also tested,
although they, especially PW2 and PW3, are essentially similar 
to conventional measures such as the ellipticity and asymmetry.

\begin{table}
 \tiny
 \caption{Summary of Subaru cluster data }
 \label{symbols}
 \begin{tabular}{@{}lcccccc}
  \hline
  Cluster & z &   filter$^a$  & exptime$^b$  & fov    \\
        & (redshift) &    & (sec)    & (arcmin)   \\
  \hline
  \hline
\scriptsize
3C295 & 0.46 & R & 960 & 26.3x33.2 \\
A115 & 0.1971 & I+ & 3600 & 35.1x27.7 \\
A1201 & 0.1688 & I+ & 2160 & 35.1x26.5 \\
A1204 & 0.1904 & I+ & 2160 & 35.3x28.4 \\
A1413 & 0.1413 & R & 4800 & 35.1x28.1 \\
A1423 & 0.213 & I+ & 3840 & 35.1x28.7 \\
A1682 & 0.226 & I+ & 2160 & 34.8x29.2 \\
A1689 & 0.184 & I+ & 2160 & 35.9x28.5 \\
A1758 & 0.28 & R & 2880 & 34.9x27.2 \\
A1763 & 0.2279 & I+ & 2160 & 35.4x29.0 \\
A1835 & 0.258 & I+ & 2400 & 34.7x25.8 \\
A1914 & 0.1712 & R & 2880 & 33.5x27.6 \\
A2034 & 0.11 & R & 2880 & 32.3x26.6 \\
A2069 & 0.1145 & R & 900 & 26.4x28.7 \\
A2111 & 0.211 & I+ & 2880 & 34.8x29.2 \\
A2204 & 0.152 & R & 540 & 28.0x28.1 \\
A2218 & 0.171 & I+ & 1500 & 36.0x28.6 \\
A2219 & 0.2281 & R & 1440 & 31.9x28.0 \\
A2255 & 0.0798 & R & 2520 & 34.5x28.8 \\
A2259 & 0.164 & I+ & 2160 & 35.3x26.9 \\
A2261 & 0.224 & R & 1620 & 26.2x31.6 \\
A2319 & 0.0564 & R & 2835 & 34.2x26.8 \\
A2390 & 0.233 & R & 1620 & 26.8x35.5 \\
A2409 & 0.1470 & I+ & 2160 & 35.3x29.3 \\
A2552 & 0.299 & R & 1680 & 26.2x34.2 \\
A267 & 0.23 & I+ & 2640 & 32.6x25.6 \\
A370 & 0.357 & I+ & 1800 & 35.0x28.9 \\
A520 & 0.203 & I+ & 1440 & 33.9x27.5 \\
A586 & 0.171 & I+ & 2100 & 35.1x26.4 \\
A611 & 0.288 & I+ & 2100 & 34.8x27.8 \\
A68 & 0.2546 & I+ & 2400 & 34.5x27.0 \\
A697 & 0.282 & I+ & 2400 & 34.0x27.3 \\
A750 & 0.163 & I+ & 1680 & 34.6x26.2 \\
A754 & 0.0528 & R & 2880 & 33.5x26.0 \\
A773 & 0.217 & I+ & 4320 & 35.3x29.4 \\
A781 & 0.2984 & I+ & 2160 & 35.3x29.0 \\
A963 & 0.206 & R & 3240 & 26.6x26.4 \\
Hercules & 0.154 & I+ & 2160 & 35.3x28.5 \\
CL0016 & 0.541 & I+ & 3600 & 33.8x25.4 \\
MS0451 & 0.54 & I+ & 2280 & 28.6x35.1 \\
MS1054 & 0.83 & Z+ & 3600 & 34.5x27.9 \\
MS1359 & 0.328 & I+ & 1800 & 35.4x15.5 \\
MS2053 & 0.583 & I+ & 3600 & 35.4x26.1 \\
RXJ0152 & 0.835 & Z+ & 5220 & 35.3x27.4 \\
RXJ0848B & 0.57 & I+ & 2160 & 35.0x26.6 \\
RXJ0848$\dagger$ & 1.27 & Z+ & 4080 & 34.9x29.3 \\
RXJ0849$\dagger$ & 1.26 & Z+ & 4080 & 34.9x29.3 \\
RXJ0910$\dagger$ & 1.106 & Z+ & 1800 & 34.5x27.5 \\
RXJ1054 & 1.134 & Z+ & 3360 & 34.7x27.4 \\
RXJ1252 & 1.235 & Z+ & 3300 & 34.9x28.1 \\
RXJ1347 & 0.451 & I & 1800 & 29.2x28.4 \\
RXJ1532 & 0.3615 & I & 2160 & 28.4x32.4 \\
RXJ1716 & 0.813 & Z+ & 3040 & 34.8x27.9 \\
RXJ1720 & 0.164 & R & 1440 & 26.8x32.9 \\
RXJ2129 & 0.235 & R & 1620 & 28.6x27.0 \\
RXJ2228 & 0.412 & I & 1620 & 32.3x27.9 \\
WGA1226 & 0.89 & Z+ & 1080 & 27.9x33.2 \\
ZWCL0024 & 0.39 & R & 5280 & 35.9x26.2 \\
ZWCL1883 & 0.194 & I+ & 2100 & 34.6x28.7 \\
ZWCL1953 & 0.3737 & I+ & 450 & 33.5x26.8 \\
ZWCL2661 & 0.3825 & R & 2880 & 28.5x26.8 \\
ZWCL2701 & 0.214 & I+ & 2160 & 35.3x28.5 \\
ZWCL3146 & 0.2906 & I+ & 1440 & 34.8x17.7 \\
ZWCL3959 & 0.3518 & I & 2160 & 34.2x29.0 \\
ZWCL5247 & 0.229 & I+ & 2160 & 35.3x29.2 \\
ZWCL7160 & 0.2578 & I+ & 3600 & 35.6x28.2 \\
  \hline
  \hline
 \end{tabular}
\\
a: detection filter, filters marked with '+' are SDSS filters, without '+' are Johnson-Cousins filters;  b: effective exposure time; $\dagger$: clusters with no obvious BCGs (excluded from the analyses related to BCGs) 

\end{table}

\section[]{Systematics and  Deblending}

\subsection[]{Systematics}

One of the important, yet unfortunately 
often lightly treated, 
problems  
associated with
comparison of  complex morphological characteristics of astronomical objects,
galaxies or clusters, 
is the  possible systematics introduced by
various data quality, exposure times and object redshifts.
Depending on the sensitivity of measures of characteristics,
some susceptible measures
may be seriously affected by these systematics, producing misleading
results.

Unfortunately, investigating the systematics on the
complex characteristics is not an easy task.
To investigate the  systematic effect of, for example, various
exposure times,
one of the  standard approaches is to
simulate an image with a given exposure time
by using an exposure-time-scaled and noise-added model image.
We need to approximate the various model characteristics 
to the complicated characteristics
of a real object. Unfortunately, those characteristics of the real objects are often what we want to
investigate, and thus assuming what we want to measure is an almost impossible task.

Meanwhile,
if we use real data, instead of a model,
we will not have this problem.
However, the standard  simple `rescaling and adding-noise' process 
to simulate a shorter exposure time will
produce an image containing
an excessive amount of
Poisson noise for a given exposure time
because of the intrinsic noise already presented in the
original data,
thus lead us to underestimate the data quality.
This intrinsic noise is difficult to be removed
even if we sacrifice the fine spatial details of
an object by smoothing, because these smoothing will
introduce yet another noise by correlating noises.

Similarly, to investigate the effect of dimming and
smaller angular size caused by higher  redshifts,
in addition to the effect of the rest waveband shift,
simple rescaling and rebinning
of the real data  will not work,
because these  manipulations will again produce the
incorrect amount of noise.

To circumvent most of these challenging problems,
\citet{hashimoto2007rqm}
developed a very useful simulating technique
employing
a series of `adaptive scalings'  accompanied by a noise adding
process applied to the real images.
This technique, which  can be used for 
all kind of imaging data, including optical, NIR, and X-ray images,
allows us to simulated an image of desired  
exposure time and redshift with correct signal-to-noise ratio.
Here we briefly describe the method, but please see Hashimoto et al. 2007 for further discussion and details.

To simulate data with integration time t1, an original unsmoothed image
(containing the background) taken with original
integration time t0 was first rescaled by a factor R$_0$/(1-R$_0$), instead of simple R$_0$,  where R$_0$=t1/t0, t0$>$t1.
That is, an intermediate scaled image I$_1$ was created from the original
unsmoothed image I$_0$ by:
\begin{eqnarray}
  I_1   &=& I_0\frac{R_0}{(1-R_0)}.     
\end{eqnarray}

Poisson noise was then added to this rescaled image by taking each pixel value as the
mean for a Poisson distribution and then randomly selecting a new pixel value from
that distribution. This image was then rescaled again by a factor (1-R$_0$)
to produce an image whose {\it signal} is scaled by R$_0$ relative
to the original image, but
its {\it noise} is approximately scaled by $\sqrt{R_0}$,
assuming that the intrinsic noise initially present in the
real data is Poissonian.

Similarly, to simulate
the dimming effect by the redshift,
an intermediate scaled image I$_1$
is created from the background subtracted image
I$_0$ by a pixel-to-pixel manipulation:

\begin{eqnarray}
  I_{1}(x,y) &=& \frac{I_{0}(x,y)^2R_1^2}{[I_{0}(x,y)R_1+B-R_1^2(I_{0}(x,y)+B)]} \\
  where & &\nonumber \\
  R_1&=&[(1+z0)/(1+z1)]^4 
\end{eqnarray}
where z0 and z1 are the original redshift and the  new redshift of the object,
respectively,
and B is the background.

Finally, to  simulate
the angular-size change due to the redshift difference
between z0 and z1, the original image will be rebinned
by a factor R$_2$, [i.e. R$_2$= (angular-size at z0)/(angular-size at z1)] , then intermediate scaled image will be
created by  rescaling the rebinned image by a factor
1/(R$^2_2$-1),
before the addition of the  Poisson noise.
For the simulation with `increased' exposure time,
this factor can be changed to
R$_3$/(R$_2^2$-R$_3$)
where  R$_3$ = t2/t0, t2$>$t0, where  t2 is the
increased exposure time, and t0 is the original integration time,
and (R$_2^2$ -R$_3$) $>$ 0.
\begin{figure}
 \center{
 {\includegraphics[height=7cm,width=5cm,clip,angle=-90]{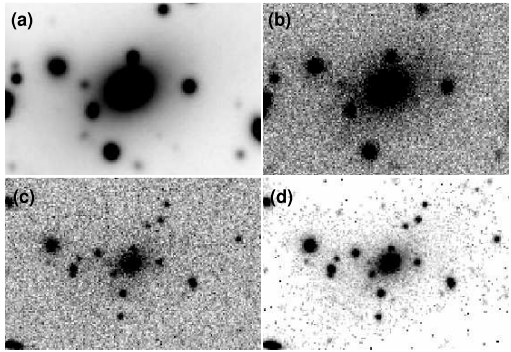}}
 }
 \caption{
Simulating an image with desired exposure time and redshift using the real data:
Even simulating an image with prolonged exposure time is possible
with our adaptive scaling method.
Here,
optical R band images,
taken with  Subaru Suprime-Cam,
around the center
of an example cluster (Abell 2219) are shown.
Images with original and modified exposure time and redshift
are presented with
north is up and east is left.
(a) Original image: exptime(t)=240s, and redshift(z)=0.228,
(b) Simulated shorter exposure image with t=10s
(c) Simulated high-z  image with z=0.9, t=240s
(d) Simulated prolonged exposure at high-z  with t=1092s, z=0.9
}
\label{FigTemp}
\end{figure}

\begin{figure}
 \center{
 {\includegraphics[height=7cm,width=5cm,clip,angle=-90]{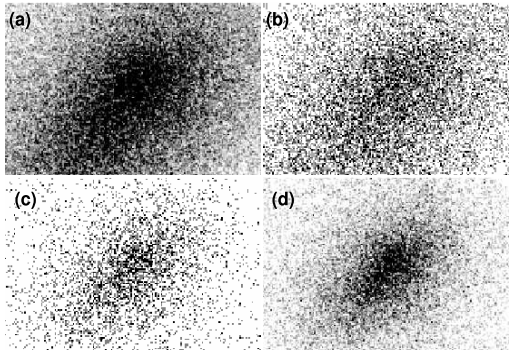}}
 }
 \caption{
Similarly with Fig. 1,
X-ray images from Chandra ACIS,
of Abell 2219 are shown
with original and modified exposure time and redshift.
North is up and east is left.
(a) Original: t=41ks, z=0.228
(b) Simulated shorter exposure: t=10ks (z=0.228)
(c) Simulated high-z image: z=0.9 (t=41ks)
(d) Prolonged exposure at high-z: t=188ks, z=0.9
}
\label{FigTemp}
\end{figure}

Although we suspected that our 
various measurements
were quite robust, as a precaution
we investigated  the possible systematics
on these measures
introduced by  various  exposure times and redshifts, using
our scaling technique described above.

In Fig. 1, 
we demonstrate
our technique of
simulating desired exposure time and redshift using the real
optical image around  the BCG at the cluster center taken with  Subaru Suprime-Cam.
Original and modified exposure time and redshift  of an example
cluster (Abell 2219) are shown with north  up and east left,
where (a) original image: exptime(t)=240s, and redshift(z)=0.228 ($\sim$ 20''$\times$35'' area is displayed here),
(b) simulated shorter exposure image with t=10s,
(c) simulated high-z  image with z=0.9, t=240s, and
(d) simulated prolonged exposure at high-z  with t=1092s, z=0.9.

 Similarly, in Fig. 2, we
 use the real
X-ray images
 of Abell 2219
from Chandra ACIS,
and simulated various exposures and redshifts, where
(a) original image with t=41ks, z=0.228 ($\sim$ 100''$\times$175'' area is displayed here),
(b) simulated shorter exposure: t=10ks (z=0.228),
(c) simulated High-z image: z=0.9 (t=41ks), and
(d) prolonged exposure at High-z: t=188ks, z=0.9.

Using this technique,
we simulated datasets with various exposure  times
and redshifts, and measured our cluster and galaxy  parameters.
We found that
our X-ray and optical measures, 
at least of a single object, 
were
quite
robust against various exposure times and redshifts.
For more detail, please see  
\citet{hashimoto2007rqm}.
For the influence from the neighboring objects and
minimum number of pixel will be further discussed in
the section 4.2.

\subsection[]{Deblending and Minimum Number of Pixel}

Our measures proved to be relatively independent of
the direct influence from the variation of data quality.  
However, for measuring galaxy morphology, 
it can be easily speculated that a threshold related to
minimum number of pixel of galaxy should be introduced.
This threshold, however, actually consists of two types.
The first type is the threshold to deal with the the direct effect from 
the
small number of pixel (npix) of a single galaxy.
The level of the threshold of this type is dependent on the particular morphological
measure. We find that, in general, the measures such as concentration index require the largest
threshold (npix $\sim$ 100).
The second type of the threshold is to deal with the indirect influence
related to the `deblending' of multiple galaxies.
This type of threshold is comparatively 
less recognized, therefore further explanation is necessary.

Suppose that we have an imaginary  pair of galaxies in simulation
whose intrinsic characteristics (such as  physical size, absolute magnitude, or intrinsic colour) remain constant while we move the pair across a range of redshifts during the simulation. 
Apart from the effect of `k correction', 
as the redshift of the pair changes,
the apparent brightness (relative to the background noise) and  
the angular size of each individual galaxy
also changes.  
In addition,
the angular $separation$ 
of the pair becomes smaller 
and this small angular separation (relative to the size
of resolution elements or seeings) and small apparent size of each galaxy
will `smooth' the light distribution and it 
reduces the relative `contrast' of two galaxies.
Similarly, any type of variations in the data quality, such as
lower spatial resolution or lower signal-to-noise, can cause similar
changes in 
those four characteristics (i.e. the apparent brightness and angular size of
each galaxy, and the angular separation and relative contrast of the pair).

Unfortunately,
our ability to separate the pair galaxies from each other,
i.e. `deblending ability' in an image detection algorithm is predominantly dependent 
on these
four characteristics. 
Therefore, variations in redshifts or data quality 
will cause the variations in the `deblending'
of galaxies, and thus may indirectly affect the 
morphological measures. For example, if you have two
neighbouring `undistorted' galaxies,  and 
bring them to a high redshift or degrade the image quality,
they start to resemble a single big  morphologically `distorted' galaxy,
instead of two separate normal looking galaxies.
This effect is unfortunately not unique to our morphological
measures, but can occur at any qualitative or quantitative
investigations of morphology. 

One can improve the deblending performance by using data
of better image quality (such as higher spatial resolution and/or deeper exposure). Similarly, one can improve the performance by changing 
the deblending parameters in the detection algorithm.
However,  
this `improved' deblending will again encounter the same deblending 
problem at more subtle deblending conditions (such as the smaller apparent separation of two galaxies).   
Furthermore,  it is fundamentally not trivial, therefore somewhat arbitrary, to
determine the optimal deblending performance. Namely, it is not trivial to determine if an 
object is really composed of two galaxies or a single morphologically disturbed galaxy with two cores.   

Fortunately, we only have two deblending statuses, i.e. `deblended' or `blended',
therefore the effect of `deblending capability' should 
remain roughly constant down to a
certain `deblending limit', then it drops.
(Therefore, even if we apply a significant deblending, it will not change
our morphological measures 
much compared to the `optimal' deblending, unless we do it to
extremes, where the algorithm starts to deblend small
intra-galactic structures.)
Furthermore, this `deblending capability' can be roughly
characterized by the apparent size of object (e.g. number of pixel of 
a galaxy), because the average surface density (therefore,
the average separation) of galaxies is related to apparent magnitude of
galaxies, and this magnitude is approximately correlated to the number of pixel of galaxy. 

When we plot the mean value of each measure versus the number of
pixel of an object inside the entire image, 
one can find that the mean value is more or less
flat down to a certain number of pixel, then object consists of
smaller number of pixel starts to show much different (usually bigger) mean 
value of the morphological  measure.
The effect of the deblending on the minimum number of pixel
can be best characterized by the ellipticity or PW2/PW0,
and for the nominal setting for the deblending in your image
extracting software, the threshold should be set around npix=50.

Now, this threshold related to the deblending is bigger than 
previously explained threshold related to the single object
for the measure such as ellipticity
and PW2, but we find that it is smaller than the single object threshold for measures
such as the concentration and asymmetry. Note that
the single object threshold is, as explained earlier in the section, 
variable among different measures.
Instead of applying measure-by-measure threshold, we decided to
apply one generic threshold (npix=100) for all measures,
estimated according to the measure that requires the largest
minimum number of pixel.

\section[]{Results}

\subsection[]{Distributions of X-ray Characteristics of Clusters}

\begin{figure}
 \includegraphics[height=6cm,width=9cm,clip,angle=0]{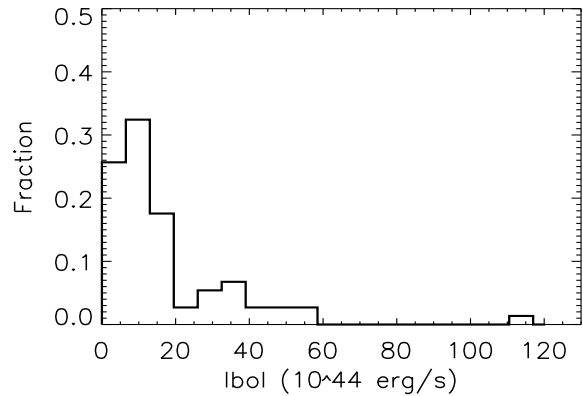}
 \caption{
Distribution of the X-ray bolometric luminosity (L$_{bol}$) for our cluster sample.
}
\label{FigTemp}
\end{figure}

Figure 3 shows the distribution of  X-ray bolometric luminosity
of our cluster sample. 
The luminosity ranges  between 1.0 $\times$ 10$^{44}$ -- 1.2 $\times$ 10$^{46}$ erg s$^{-1}$
(median 8.56 $\times$ 10$^{44}$ erg s$^{-1}$),
while
Figure 4 and 5   show distributions of cluster X-ray morphology in the
Cx (concentration) vs. Ax (asymmetry) plane, and Cx vs. Ex (ellipticity) plane,
respectively.
One sigma errors are approximately estimated from the Monte Carlo simulation.
Note that entire X-ray sample of 
\citet{hashimoto2007rqm} is plotted here. 
In Fig. 4 and 5, we can see that clusters are scattered in the morphological
planes, showing various morphological characteristics.
However, there is a strong to weak correlation between Cx and Ax, and Cx and Ex with the value of
Spearman $\rho = $-0.52 and -0.36 (significant at significance level = 1.77 $\times$ 10$^{-9}$ and 4.95 $\times$ 10$^{-5}$), respectively.
These trends indicate that low concentration clusters generally show high 
degree of asymmetry, or ellipticity, illustrating the fact that there are
not many highly-extended diffuse clusters with  symmetric round profiles.
Note that Fig. 5 shows a weak correlation between Ex-Cx. We suspect that
this trend is due to the intrinsic nature of cluster morphology, and not
due to the projection effect, because our concentration index is 
quite robust against the variation caused by the projection effect.
We will further investigate this issue in detail in Sec. 5.3.

\begin{figure}
 \resizebox{\hsize}{!}{\includegraphics[height=3cm,width=2cm,clip,angle=90]{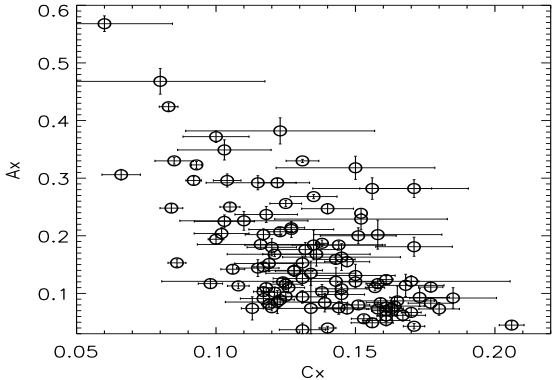}}
 \caption{
The distribution of Concentration and Asymmetry of cluster X-ray profile
for our cluster sample.
}
\label{FigTemp}
\end{figure}

\begin{figure}
 \resizebox{\hsize}{!}{\includegraphics[height=3cm,width=2cm,clip,angle=90]{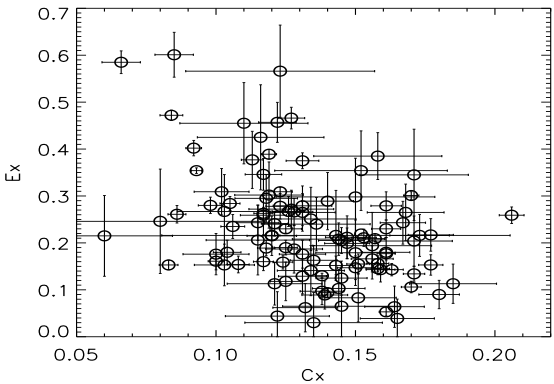}}
 \caption{
The distribution of
Concentration and Ellipticity of cluster X-ray profile
for our cluster sample.
}
\label{FigTemp}
\end{figure}

\subsection[]{Distributions of Optical Characteristics of BCGs}

\begin{figure}
 \resizebox{\hsize}{!}{\includegraphics[height=3cm,width=2cm,clip,angle=90]{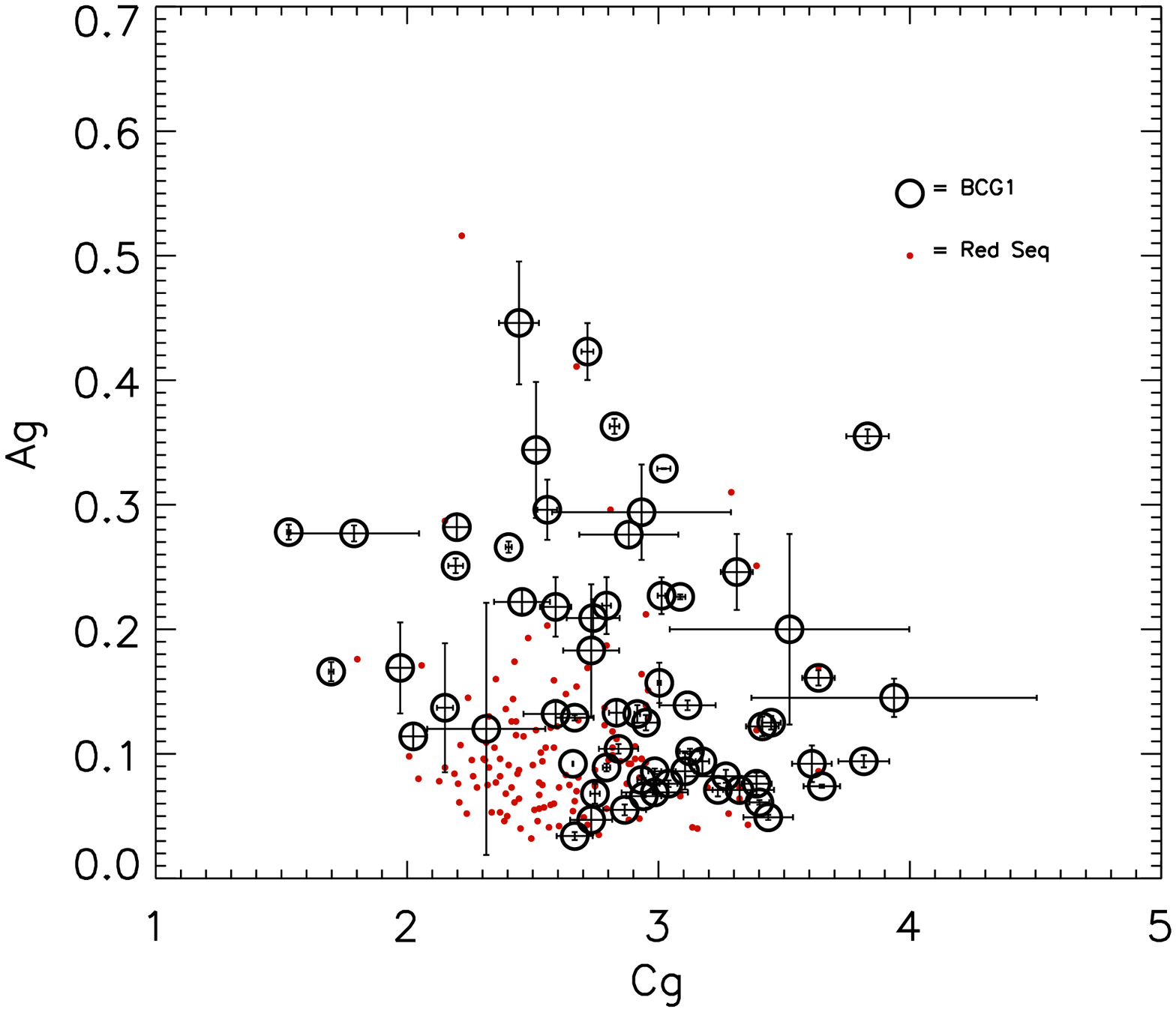}}
 \resizebox{\hsize}{!}{\includegraphics[height=3cm,width=2cm,clip,angle=90]{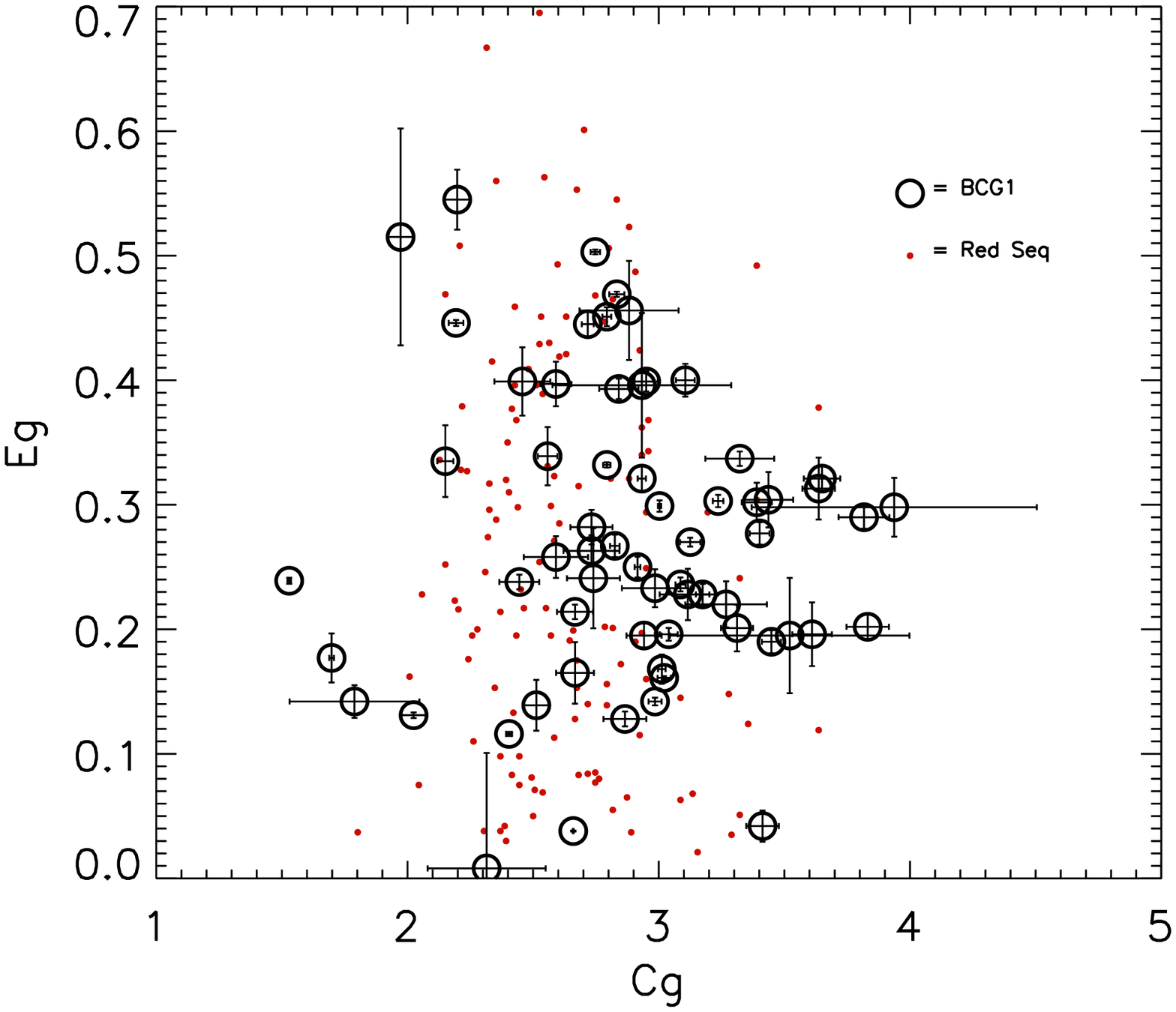}}
 \caption{
Distribution of BCG (marked by open circles) morphologies 
in the A$_g$-C$_g$ (top) and E$_g$-C$_g$ (bottom) planes.
For comparison, the randomly selected red sequence galaxies
are plotted (red dots).
}
\label{fig:CgvsMorphBCG01}
\end{figure}

\begin{figure}
 \resizebox{\hsize}{!}{\includegraphics[height=3cm,width=2cm,clip,angle=90]{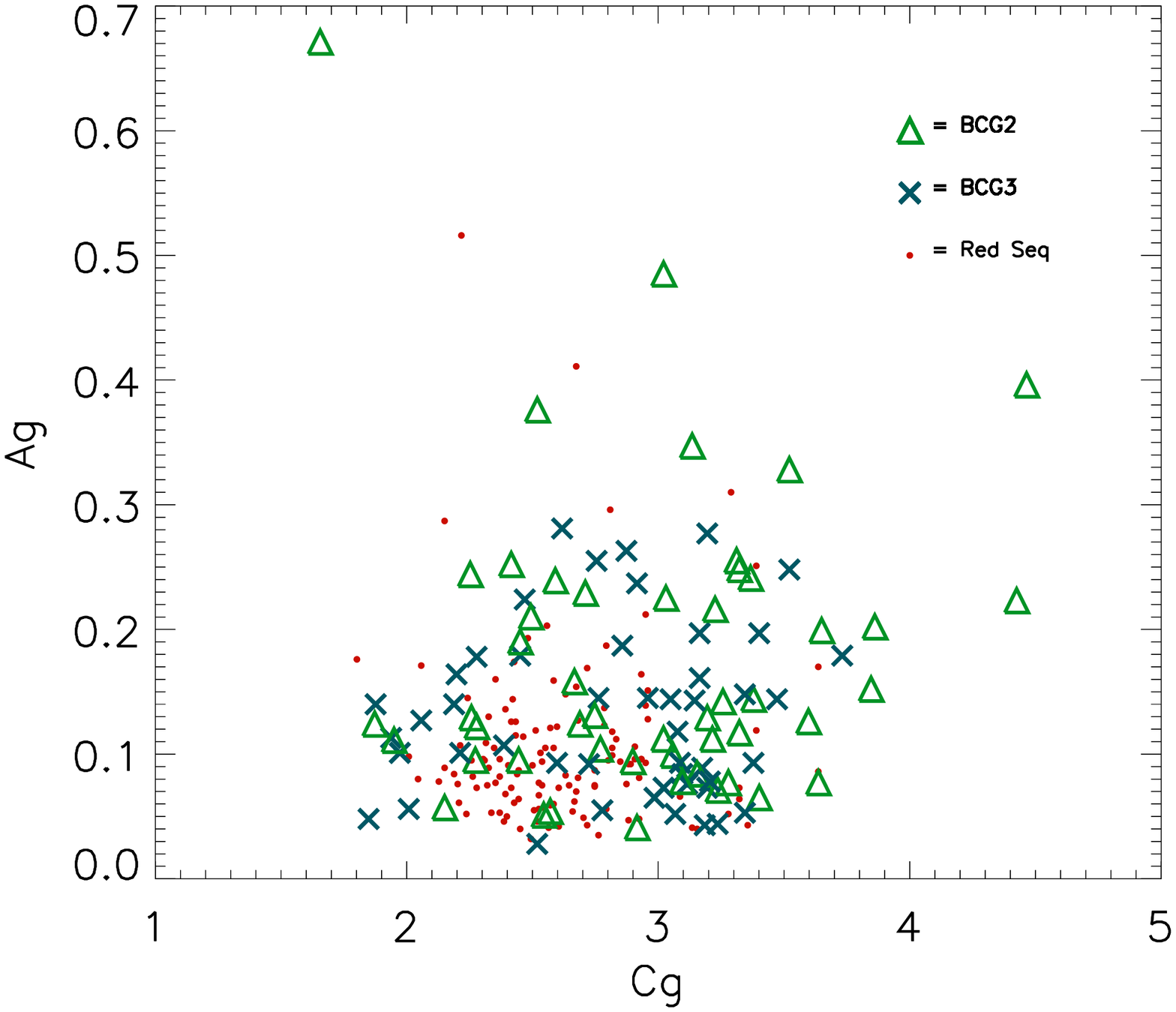}}
 \resizebox{\hsize}{!}{\includegraphics[height=3cm,width=2cm,clip,angle=90]{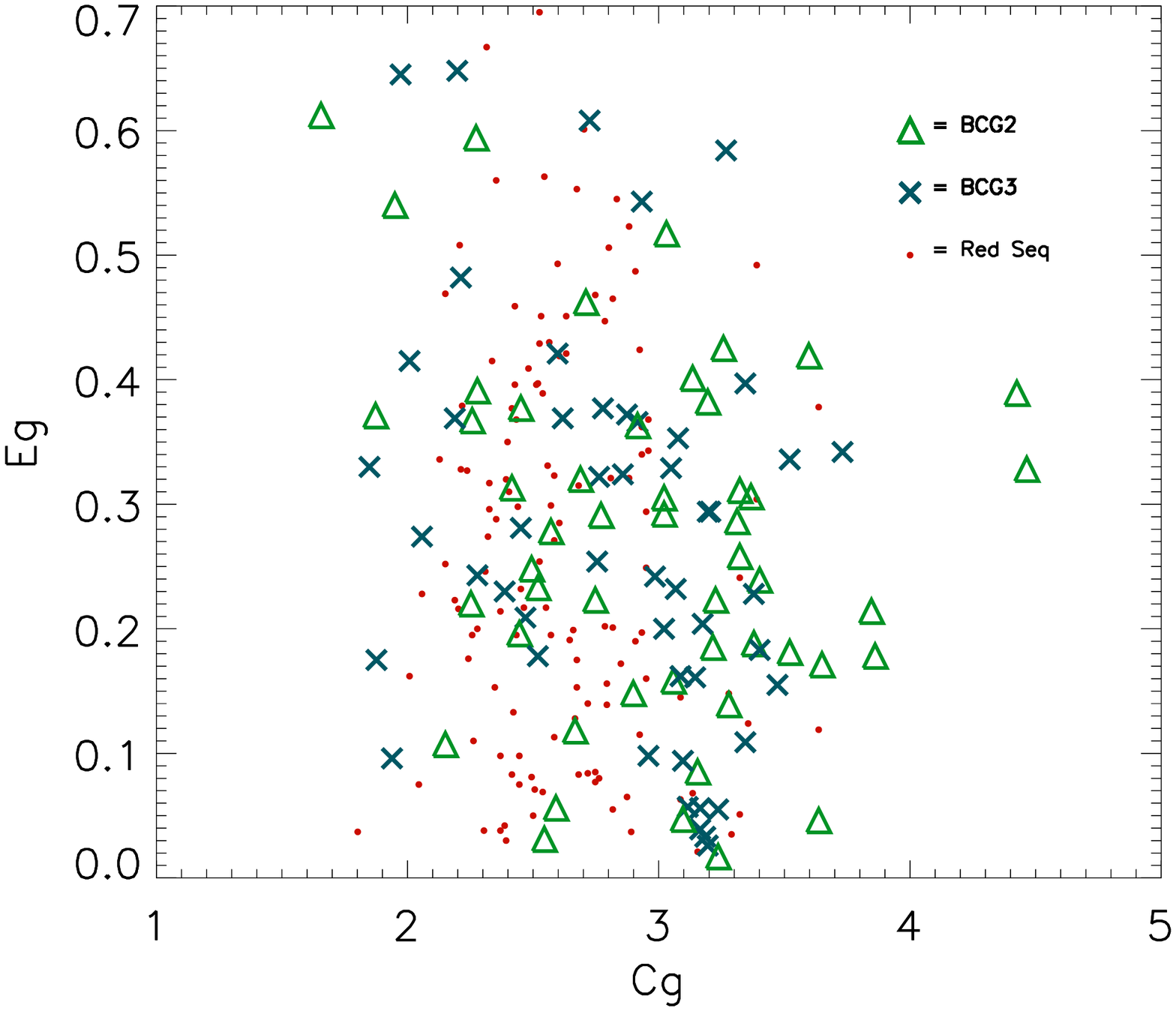}}
 \caption{
Distributions of morphologies of 
 the second brightest galaxy (BCG2; triangle), and the third brightest
galaxy (BCG3; x ) inside the projected radius of 0.5 Mpc from the X-ray center
are shown.
Similarly to Figure \ref{fig:CgvsMorphBCG01}, the red sequence galaxies are also plotted for comparison.
}
\label{fig:CgvsMorphforBCG02}
\end{figure}

Figure \ref{fig:CgvsMorphBCG01} shows the distributions of BCG morphology of our sample 
clusters in the A$_g$ (asymmetry) -C$_g$ (concentration) (top panel) 
plane, and E$_g$ (ellipticity) -C$_g$ (bottom panel) plane in the 
detection bands. 
In Figure \ref{fig:CgvsMorphBCG01}, BCGs (marked by open circles) of our 
sample clusters are showing a wide variety of
morphology in the A$_g$-C$_g$ and E$_g$-C$_g$ planes.
One sigma errors are approximately estimated from Monte Carlo simulations.
For comparison, the red sequence galaxies 
are also plotted (red dots) in Figure \ref{fig:CgvsMorphBCG01}.
For brevity, and so the number galaxies are comparable to BCGs, 
the red sequence galaxies are randomly selected from several randomly selected clusters.
The red sequence galaxies are selected  
using the color magnitude diagram, as galaxies 
brighter than the BCG magnitude + 3 and 
$\pm$ 0.1$\times$(B-R) or(V-I+) 
at either side of the semi-manually fitted red sequence
line.
On contrary to the naive
expectation that BCGs may resemble  early type galaxies,
BCGs seem to show a wider variety of morphology,
particularly in C$_g$ or A$_g$ space.
A K-S test shows that probability that distributions of BCGs and red sequence galaxies are drawn from 
the same parent distribution is 
4.14 $\times$ 10$^{-5}$ and 7.59 $\times$ 10$^{-5}$, 
for C$_g$ and A$_g$, respectively.

In Figure \ref{fig:CgvsMorphforBCG02}, the morphological distributions 
of the second brightest galaxy (BCG2), and the third brightest
galaxy (BCG3) inside the projected radius of 0.5 Mpc from the X-ray center
are shown.
BCG2 and BCG3 are then visually inspected
and the assignments are adjusted, if needed, 
based on  
their colours and the distance to the X-ray center, as well as the redshifts if available in the literature. 
Similarly to Figure \ref{fig:CgvsMorphBCG01}, the red sequence galaxies are also plotted for comparison.
BCG2 galaxies seem to show a similar morphological distribution with 
BCGs, showing a wider morphological variation than the red sequence
galaxies.
A K-S test shows that probability that distributions of BCG02s and red sequence galaxies are drawn from 
the same parent distribution is 2.61 $\times$ 10$^{-6}$ and 1.25 $\times$ 10$^{-5}$
for C$_g$ and A$_g$, respectively. 
Meanwhile,
BCG3 galaxies seem to show more or less similar morphology with BCG2, but
show a hint of slightly less clear segregation in Ag from the red sequence galaxies.
A K-S test shows that probability that distributions of BCG03s and red 
sequence galaxies are drawn from the same parent distribution is 
2.61 $\times$ 10$^{-6}$ for C$_g$, and 4.48 $\times$ 10$^{-4}$ for A$_g$. 
 The result seems to suggest that
 there is a continuous variation of morphology
 between BCG, BCG2, and  BCG3,
 rather than  a clear sharp separation of morphological characteristics 
 between the BCG and the rest of
 the bright galaxies.

\begin{figure}
 \resizebox{\hsize}{!}{\includegraphics[height=3cm,width=2cm,clip,angle=90]{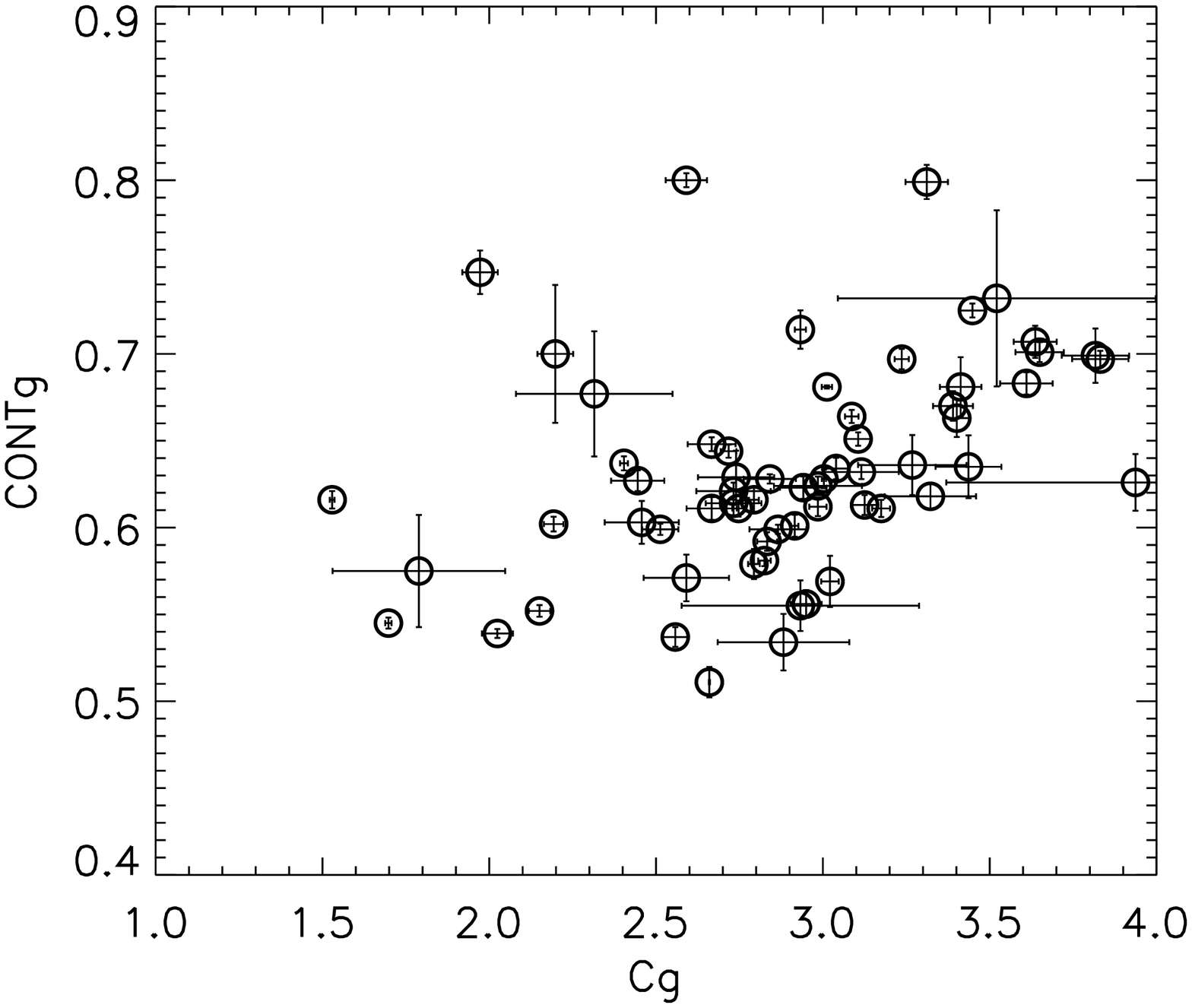}}
 \resizebox{\hsize}{!}{\includegraphics[height=3cm,width=2cm,clip,angle=90]{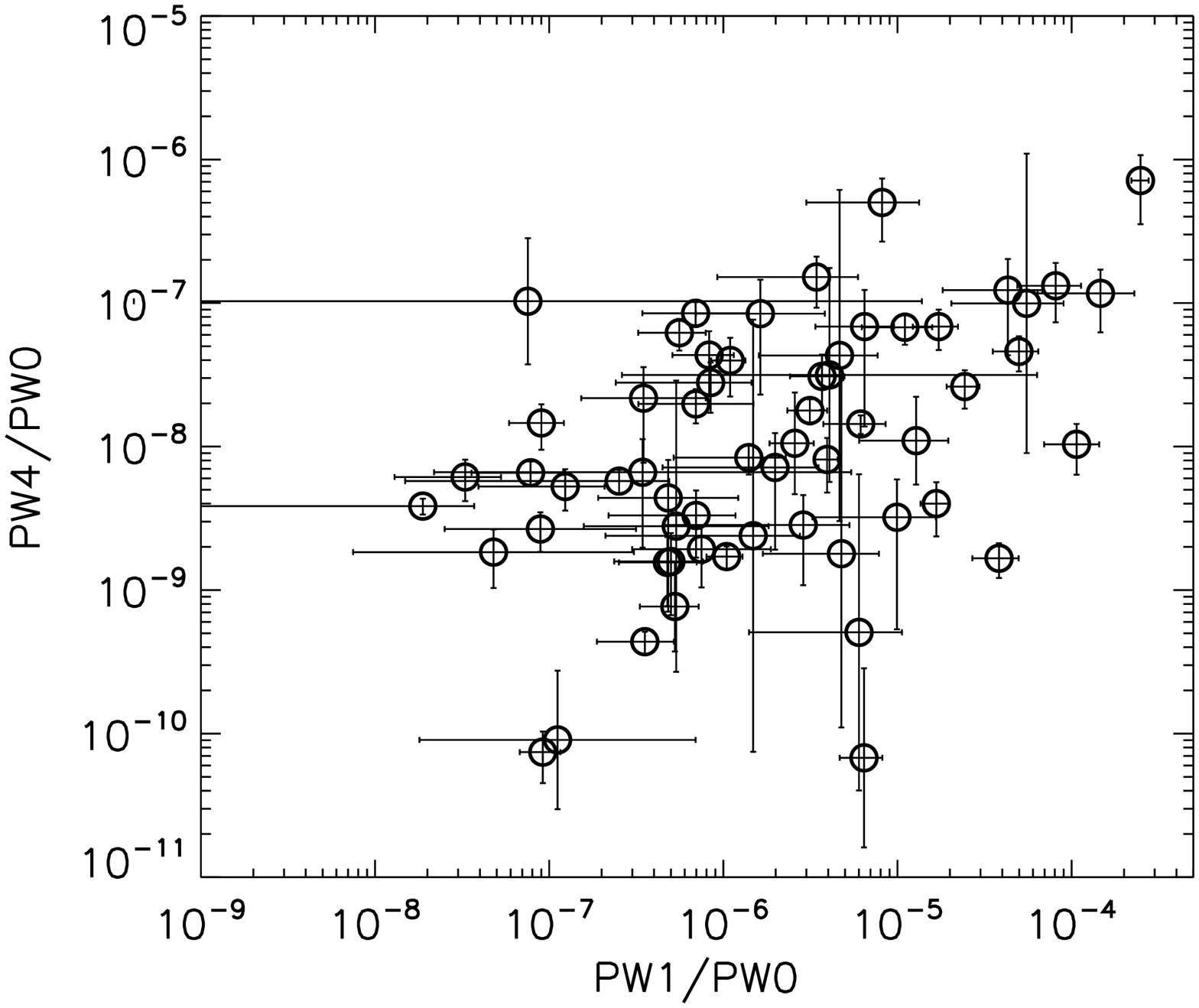}}
 \caption{
Distribution of BCGs in the planes of  further morphological measures:
Contrast vs. Ellipticity  (top panel) and PW4/PW0 vs PW1/PW0 (bottom panel). 
}
\label{fig:newmorp}
\end{figure}

Meanwhile, in Figure \ref{fig:newmorp}, the distributions of other morphological measures of 
BCGs, Contrast vs. Ellipticity  (top panel) and PW4/PW0 vs. PW1/PW0 (bottom panel) are plotted.
Please note that dipole moment PW2 and PW3 are omitted, because they
are very similar to ellipticity and asymmetry, respectively, by definition.

\subsection[]{Brightest Cluster Galaxy and X-ray Characteristics of the Host Cluster }

\begin{figure}
 \resizebox{\hsize}{!}{\includegraphics[height=3cm,width=2cm,clip,angle=90]{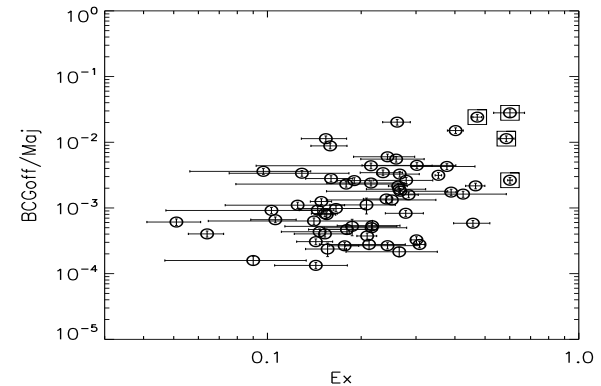}}
 \caption{
The BCG offset from the cluster center versus the ellipticity of cluster X-ray profile. The BCG offset is expressed in the unit of cluster major axis. 
The squares denote obvious double clusters. 
}
\label{fig:ExvsBCGoffset}
\end{figure}

\begin{figure}
 \includegraphics[height=9cm,width=6cm,clip,angle=90]{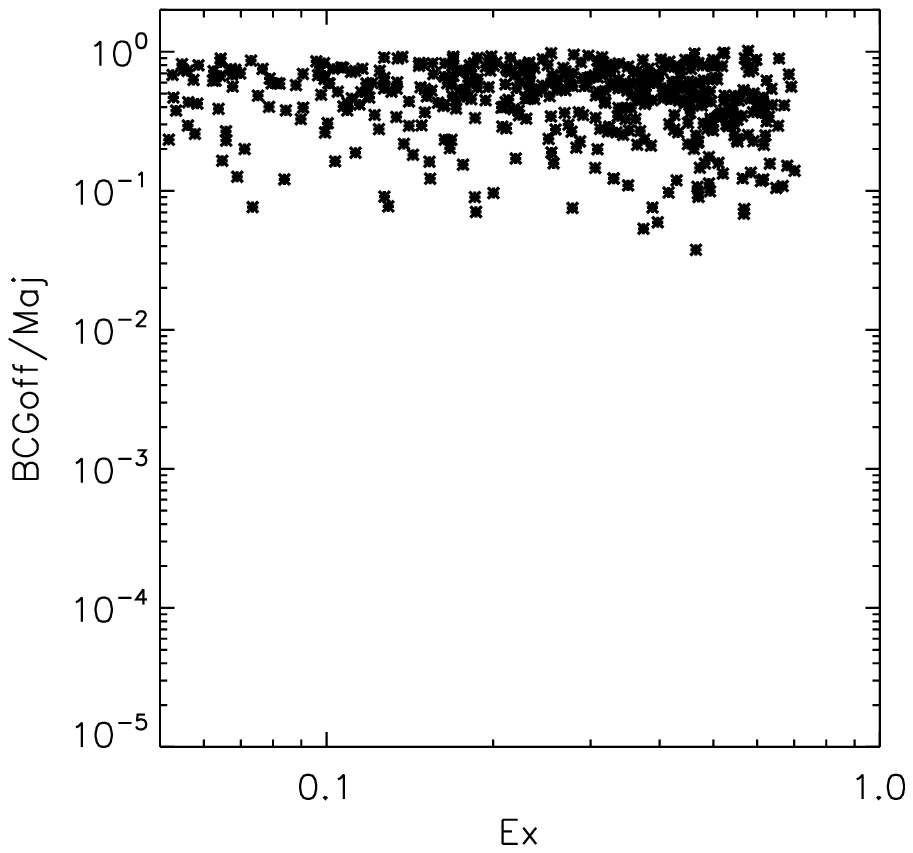}  
 \includegraphics[height=9cm,width=6cm,clip,angle=90]{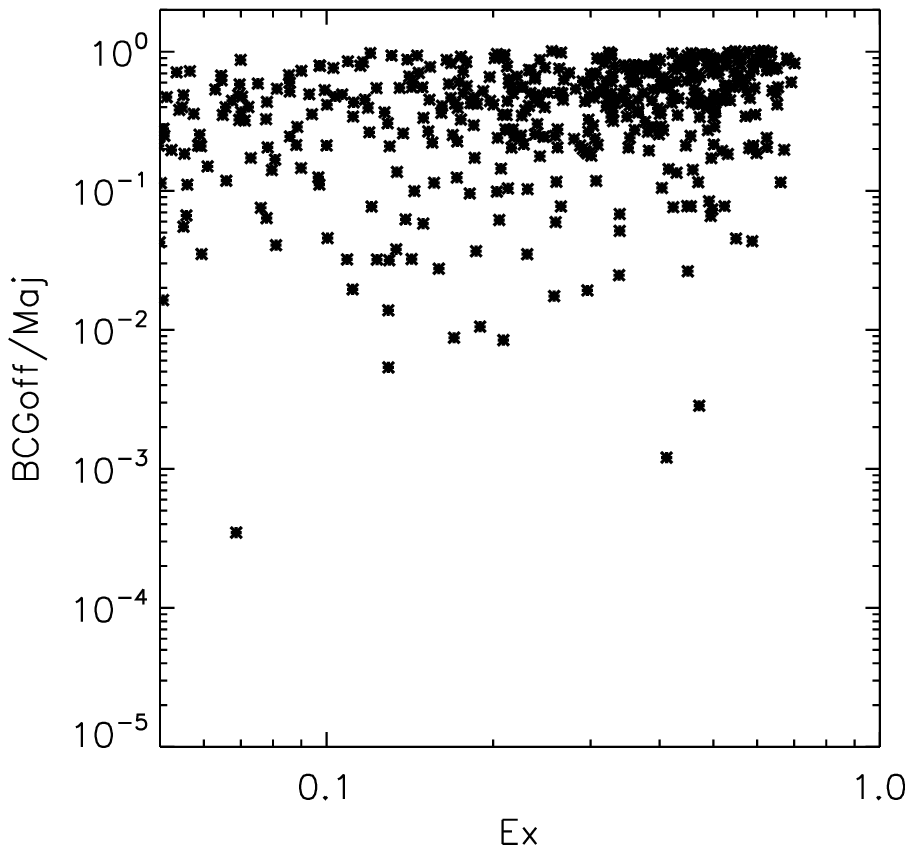}  
 \caption{
Simulated projection effect on the BCG offset vs X-ray ellipticity.
Cluster shape is assumed to be ellipsoidal with NFW profile.
The original ellipticity of the cluster (i.e. the maximum projected ellipticity
of the ellipsoid ) is chosen randomly between 0.05 and 0.7.
In the top panel, the position of BCG is randomly chosen inside the ellipsoid,
while in the bottom panel,
the position of BCG is chosen randomly $only$ along the major axis of the 
ellipsoid.
No trend between BCG offset and the ellipticity 
the cluster morphologies can be generated from the random projection effect 
of clusters,
even if we let 
the projection effects act on the ellipticity and BCG position 
in a correlated manner.
}
\label{fig:ExvsOffsetSim}
\end{figure}

\begin{figure}
 \resizebox{\hsize}{!}{\includegraphics[height=3cm,width=2cm,clip,angle=90]{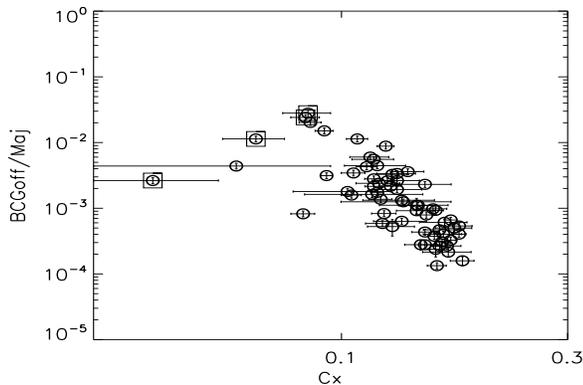}}
 \caption{
The BCG offset (in the unit of the cluster major axis) versus the concentration index of cluster X-ray profile. 
}
\label{fig:CxvsBCGoffset}
\end{figure}

\begin{figure}
 \includegraphics[height=9cm,width=6cm,clip,angle=90]{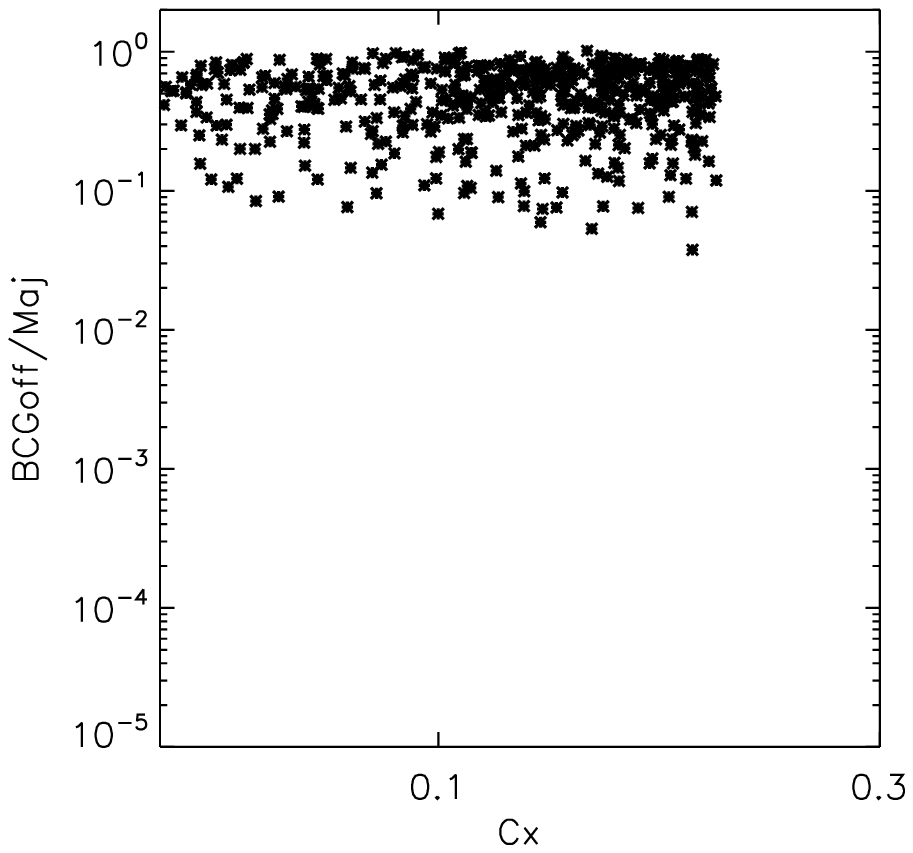}
 \includegraphics[height=9cm,width=6cm,clip,angle=90]{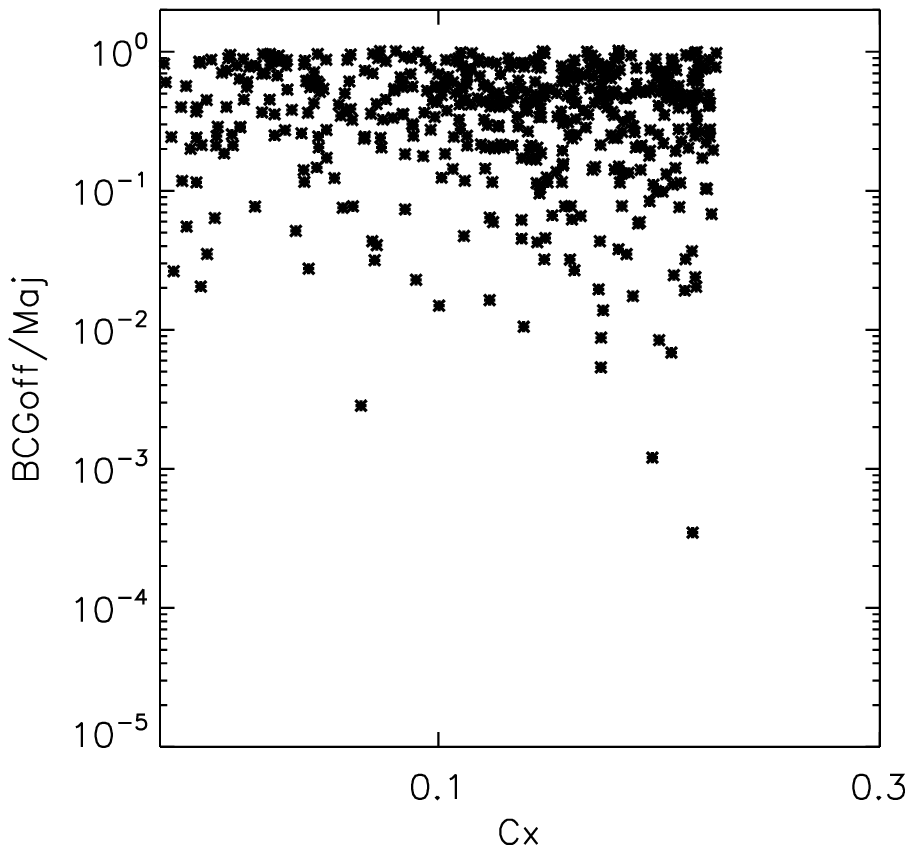}
 \caption{
Simulated projection effect on the BCG offset vs X-ray concentration.
Initial input concentration is chosen randomly between 0.05-0.2.
The position of the BCG is chosen randomly inside the ellipsoidal cluster
(top panel), and is chosen randomly $only$ along the major axis of the 
ellipsoid (bottom panel).
No trend between BCG offset and the concentration can be
generated  
from the random projection effect 
of clusters,
}
\label{fig:CxvsOffsetSim}
\end{figure}

Figure \ref{fig:ExvsBCGoffset}  
shows the distance of the BCG to the center of cluster,
expressed
in the unit of cluster major axis, plotted against the ellipticity
of cluster X-ray profile. The cluster major axis is measured from
the X-ray emission, while the BCG position is determined in the optical
detection band. The cluster center here is determined by the `4th order'
of the centroid of the X-ray emission, to minimize the influence from  faint outer X-ray structure \citep[c.f.][]{hashimoto2007rqm}.
One sigma errors on BCG offset are estimated from Monte Carlo.
It appears in Figure \ref{fig:ExvsBCGoffset}, that
there is a weak trend in such way that
clusters with high X-ray ellipticity show large BCG offsets
(Spearman $\rho$ = 0.41; significance level = 9.8 $\times$ 10$^{-4}$).

However, Figure \ref{fig:ExvsBCGoffset} should be interpreted with caution, because both the ellipticity and BCG offset can be
influenced by the cluster projection effect on the plane perpendicular to the line of sight.
Fortunately, even from a simple geometrical argument, one can estimate that the probability of creating the clusters with apparent low projected ellipticity out of high intrinsic ellipticity is rather  low, and therefore the bulk of the clusters maintains similar
or only slightly lower ellipticity than the original ellipticity.
Exactly the same geometrical argument is applicable to the offset of the BCGs.

To ensure that even this small probability will not create any artificial
trend, we performed Monte Carlo simulation.
Figure \ref{fig:ExvsOffsetSim} shows the BCG distance to the cluster center for the simulated
clusters versus projected cluster ellipticity.
Here, cluster shape is assumed to be ellipsoidal with NFW  
\citep[Navarro-Frenk-White;][]{nfw96} profile with R$_s$=0.5.
The original ellipticity of the cluster (i.e. the maximum projected ellipticity
of the ellipsoid ) is chosen randomly between 0.05 and 7,
then
the ellipsoid is randomly oriented with respect to the line of sight.
In the top panel, the initial position of BCG is randomly chosen inside the entire
ellipsoid,
while in the bottom panel, 
the position of BCG is allowed to vary randomly $only$ along the major axis of the clusters. 
In total, 500 such random clusters are generated, for each panel. 
Figure \ref{fig:ExvsOffsetSim} shows that, 
even if we let 
the projection effects act on the ellipticity and BCG position 
in a correlated manner,
the projection effect alone cannot produce the 
trend between BCG offset and the ellipticity of
the cluster X-ray morphologies.

In Figure \ref{fig:CxvsBCGoffset}, we plotted the BCG offset,
again
measured in a unit of the cluster semi-major axis, against the cluster X-ray
concentration (C$_x$) for our real sample.
The concentration, just as the ellipticity, is  expected to be sensitive to the cluster morphological distortion related to dynamical status of clusters,
yet
our concentration, unlike the ellipticity, is designed to be much more robust against projection effect,
because we define the concentration independent of the shape of 
the ellipsoid.
Figure \ref{fig:CxvsBCGoffset} shows a clear trend between the BCG offset and
C$_x$ of cluster (Spearman $\rho$ = -0.80; significance level = 3.25 $\times$ 10$^{-15}$) that is consistent with the scenario that the BCG position may be related the dynamical status of clusters 
\citep[e.g.][]{ostriker1975another,merritt1985distribution,katayama2003properties}.

In Figure \ref{fig:CxvsOffsetSim}, we further tested the projection effect on the concentration.
In Figure \ref{fig:CxvsOffsetSim}, original input concentration is chosen randomly between 0.05. and
0.2.
Figure \ref{fig:CxvsOffsetSim} shows that,
as expected from the definition of the concentration,
the concentration index hardly changes 
(except for induced small random scattering) 
by random
orientation of the clusters, and we cannot generate
any apparent correlation between BCG offset and the concentration 
by the projection effect alone, in the both cases of the random BCG position
inside everywhere in the ellipsoid (top panel), and
the random BCG position along the major axis of the ellipsoid (bottom panel).

Three clusters at the high end of cluster X-ray ellipticity (E$_x$) in 
Fig. \ref{fig:ExvsBCGoffset},
and the low end of cluster X-ray concentration (C$_x$) in 
Fig. \ref{fig:CxvsBCGoffset}, are
``double" clusters that is two clusters that appear to merge into one cluster (marked by squares in both Fig. \ref{fig:ExvsBCGoffset} and \ref{fig:CxvsBCGoffset}.
In these double clusters, in extreme case, even if we have a BCG in each of two clusters and each BCG is nicely aligned to the
center of each cluster, 
the cluster morphological measures and cluster center, based on the 
entire (double) cluster, (and choosing only one BCG per cluster),
may not be able to differentiate this `each-lobe alignment' from the whole
cluster  
alignment. Note, however, that since there is no objectively clear `boundary' between one semi-double lobe cluster and two colliding clusters,
enforcing arbitrary separation of two types of clusters may artificially 
introduce some discontinuity in the analysis, and therefore in our understanding of the cluster morphology.
Whether or not one should treat double clusters as one cluster or separating each
lobe into two clusters is not at all a trivial problem.

One more caution should be exercised in interpreting Figure \ref{fig:ExvsBCGoffset} and \ref{fig:CxvsBCGoffset}.
There is some possibility that x and y  axes in the figures 
are not completely independent. Namely, one can expect that, when cluster
morphology gets highly distorted, 
it does not often maintain the simple distorted morphology, 
and that complex morphology may lead to 
a lack of obvious center.
Even if our centering algorithm can still define the center well 
in unambiguous manner for these clusters with complex morphology, 
it may not be trivial to relate the measured
center to a
dynamically important center (e.g. center of  gravity defined by
dark matter, or center in the momentum space). 
This `uncertainty' in determining the cluster center
may indirectly affect the measurement of the BCG offset.
Note that 
this effect on the measurement of the BCG offset, if any, is expected to 
mostly increase the scattering of the measured BCG offset, and not to shift the 
offset, directly. 
However, 
if the BCG offset is intrinsically very small compared to the size of the scatter, 
this scatter in the offset may 
shift the mean offset to slightly higher value, because we have no 
negative value for the offset (i.e. the scatter {\em into} the negative value
is `folded' to positive value).  
Now, this scatter can be larger for increasing cluster X-ray distortion,
therefore if the underlying BCG offset is intrinsically small, the scattering effect
on the offset may slightly enhance the X-ray morphology vs. BCG offset trend.  
However, 
in significant fraction of clusters, where
one can define the cluster center unambiguously, even if the morphology is distorted
(e.g. nice symmetrical ellipse), the BCGs still tend to be located at a large offset from
the cluster center, and follows the overall trend in the offset related
plots without any `discontinuity' between these clusters and
other more complex looking clusters. 
Finally, even if the small part of the offset trend is indeed due to the
`scattering effect', the fact remains that both BCG offset and the cluster X-ray morphology  are comparable measures in sense of characterizing
possible dynamical status of clusters.

\begin{figure}
 \includegraphics[height=9cm,width=6cm,clip,angle=90]{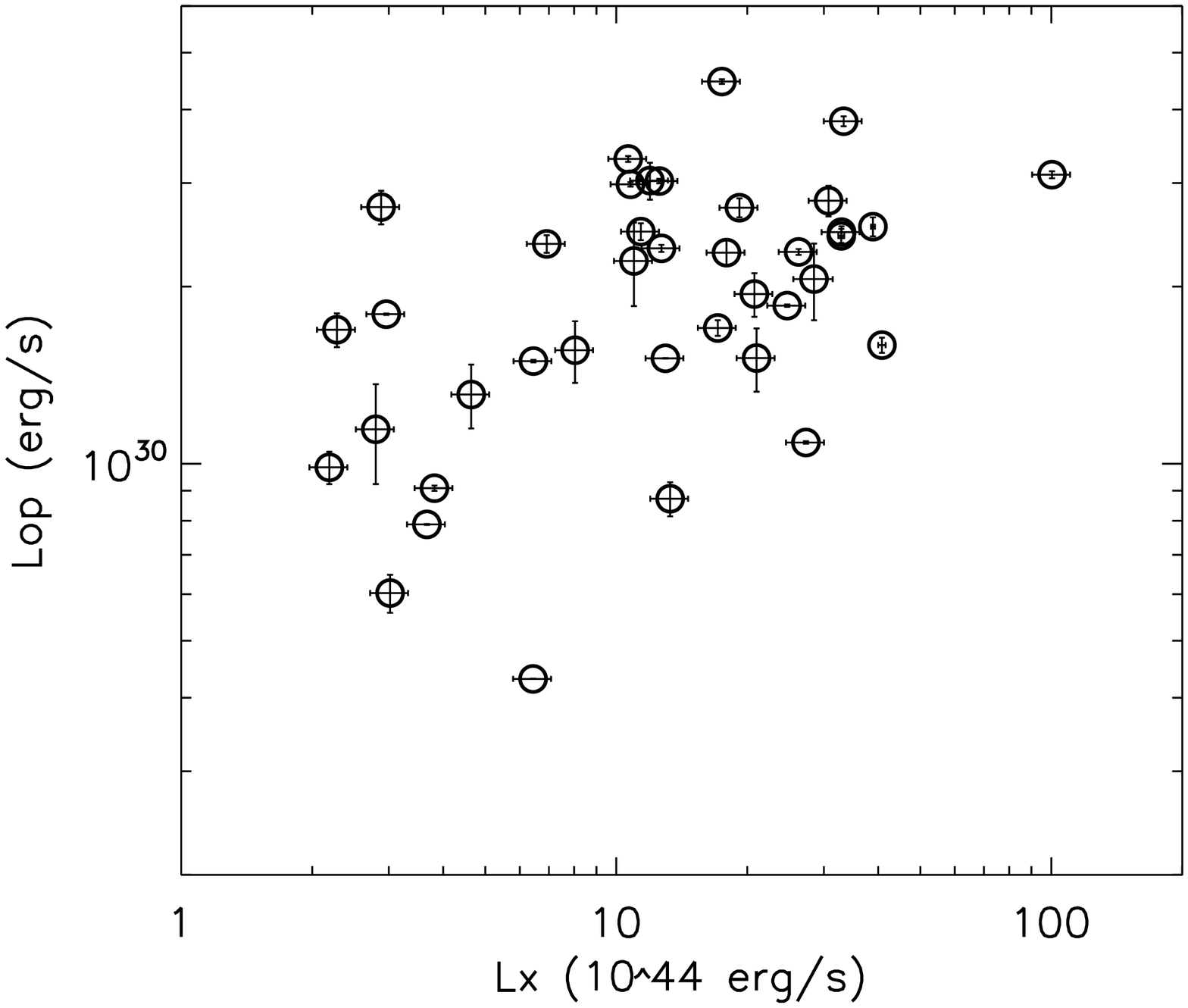}  
 \includegraphics[height=9cm,width=6cm,clip,angle=90]{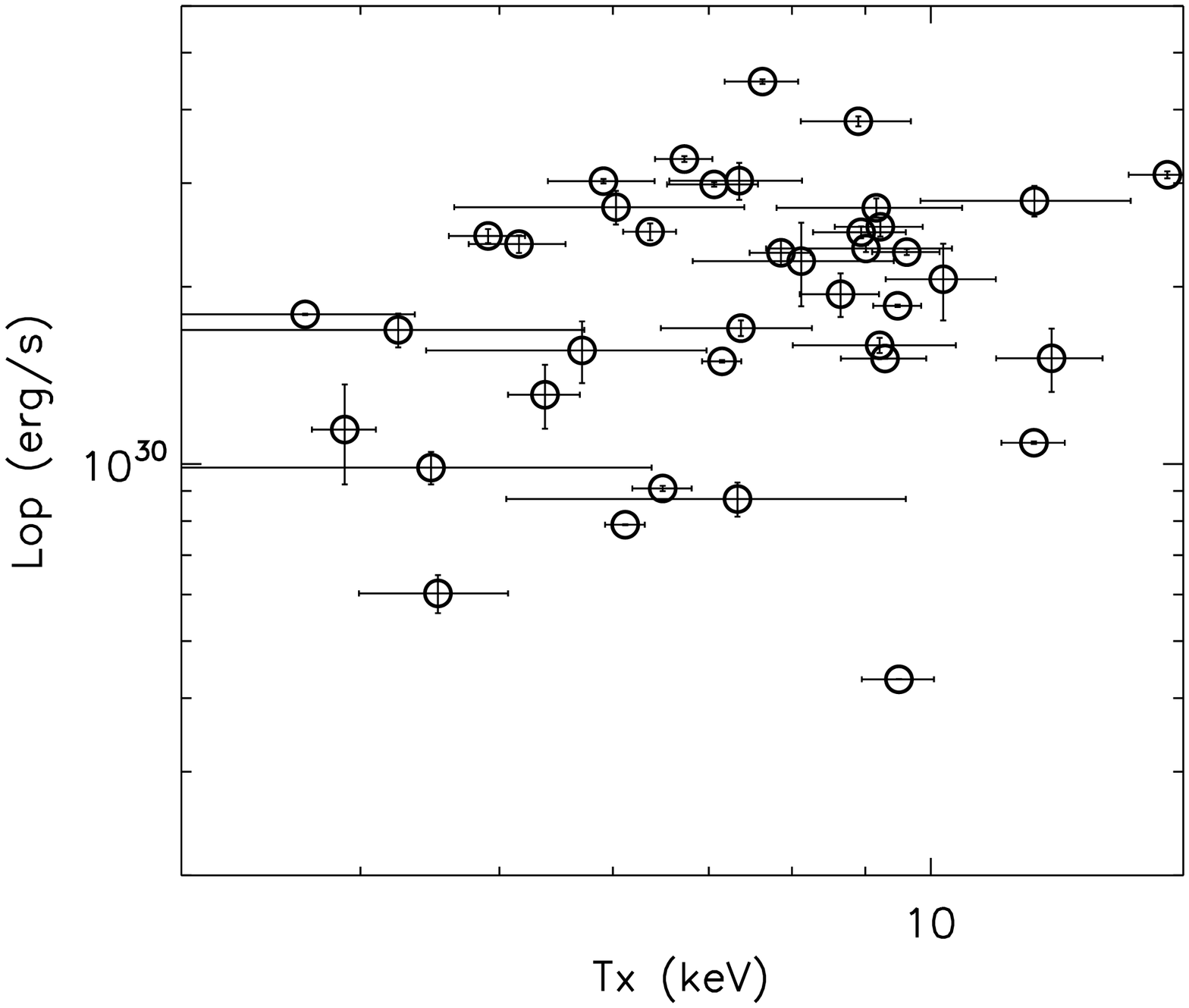}  
 \caption{
Luminosity of BCGs versus cluster X-ray bolometric luminosity (top), and versus X-ray temperature (bottom).  
}
\label{fig:LbcgvsLxTx}
\end{figure}

In Figure \ref{fig:LbcgvsLxTx}, we plot the 
optical luminosity of the BCGs 
(one from each cluster) versus  X-ray bolometric luminosity (Lx) of the host
cluster 
(top panel), and versus X-ray temperature (Tx) of the cluster (bottom panel).
The BCG luminosity is calculated  
based on the k corrected R magnitude.
Note that clusters without  R band data
are excluded in the figure, to reduce additional scatter 
associated with relative k correction error between different bands.
Lx errors are assumed to be 10\% error for clusters without estimated errors.
Both panels, in particular the top panel, show that there is a weak 
trend that high Lx or Tx clusters harbour brighter BCGs, although
the scatter is large 
(Spearman $\rho$ is 0.43 and 0.18, significance level = 6.2 $\times$ 10$^{-3}$ and 2.7 $\times$ 10$^{-1}$ respectively for the top panel and the bottom panel). 
The result is in agreement with previous works reporting
similar weak correlations between X-ray properties and optical BCG luminosity 
\citep[e.g.][]{schombert1987structure,edge1991exosat,brough2002evolution,katayama2003properties}.

\begin{figure}
 \includegraphics[height=8.3cm,width=6cm,clip,angle=90]{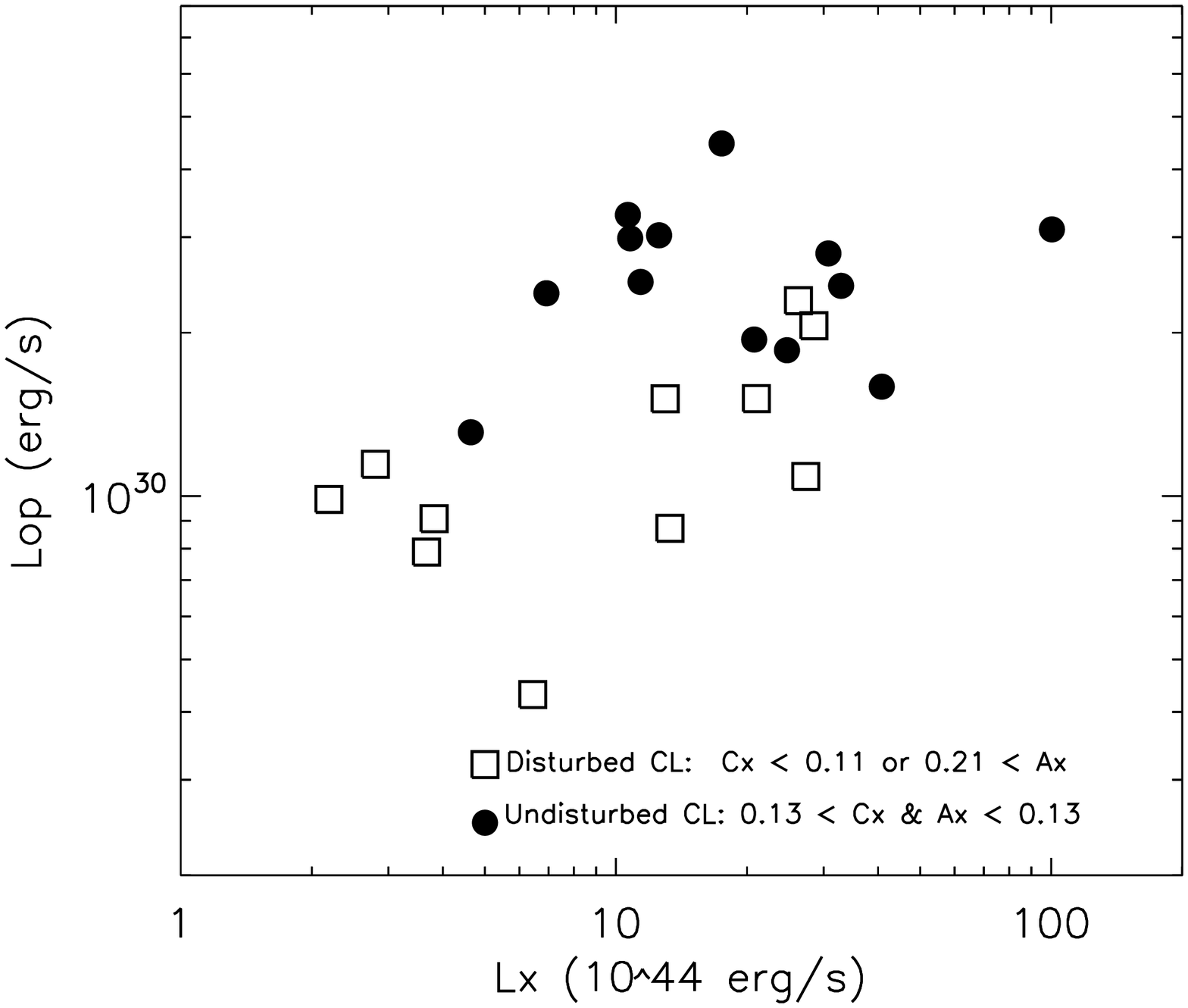}  
 \includegraphics[height=8.1cm,width=6cm,clip,angle=90]{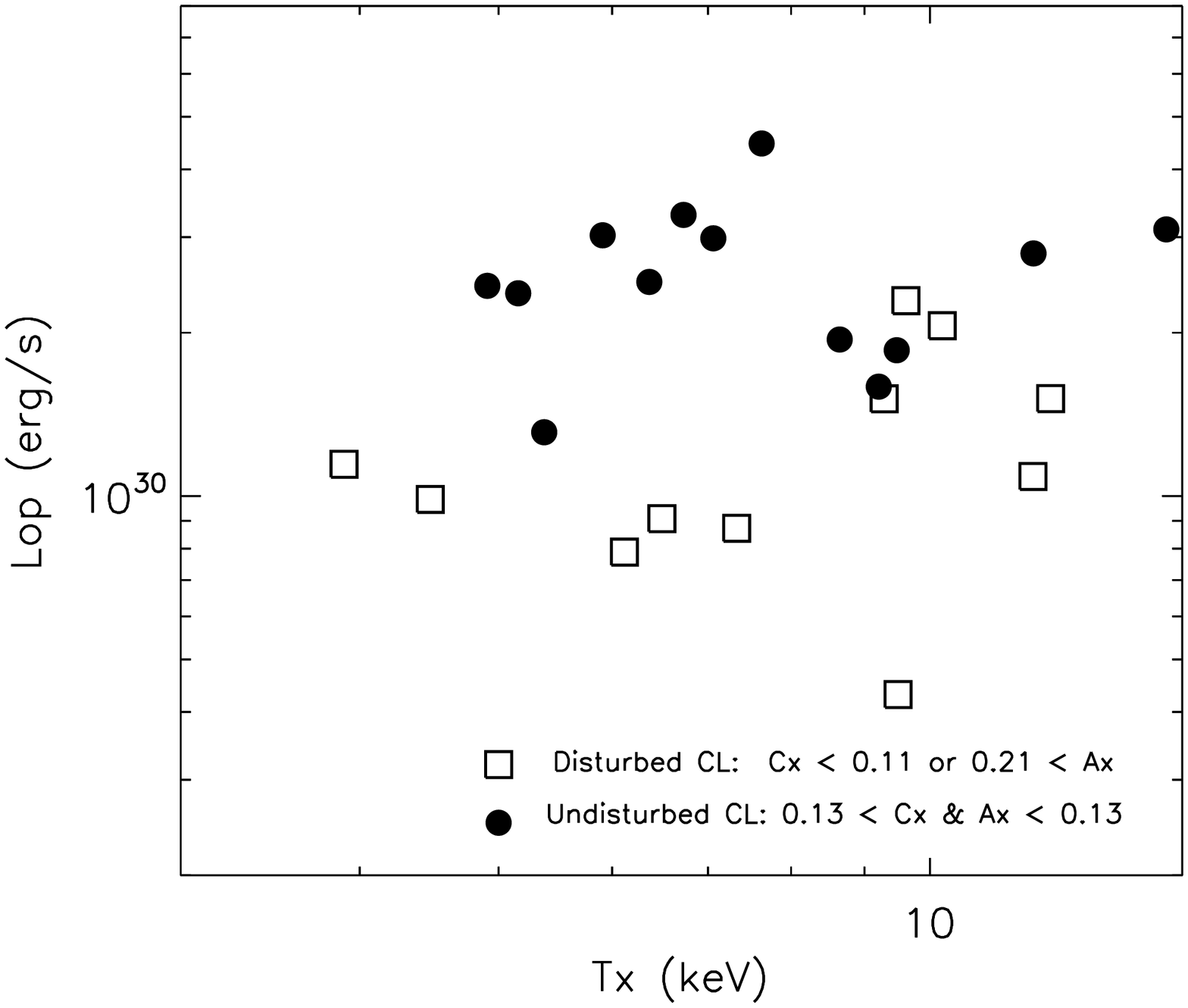}
 \caption{
Luminosity of BCG versus cluster X-ray bolometric luminosity (top), and versus  X-ray temperature (bottom)
of subsets selected based on cluster dynamical status. Solid dots represent 
clusters with highly distorted X-ray morphology 
(Cx $<$ 0.11 $or$ Ax $>$ 0.21)
while open squares represent clusters with very undisturbed morphology
(Cx $>$ 0.13 and Ax $<$ 0.13).
The plots may imply that unsettled clusters may have fainter BCGs.
}
\label{fig:LxTxvsLopwithSubset}
\end{figure}

To investigate the origin of the scattering in Figure \ref{fig:LbcgvsLxTx}, 
we show the subset of clusters according to their
X-ray morphology in 
Figure \ref{fig:LxTxvsLopwithSubset}, where the open squares represent clusters with
highly distorted X-ray morphology 
( Cx $<$ 0.11 $or$ Ax $>$ 0.21)
and the solid circles represent clusters with very undisturbed X-ray
morphology 
( Cx $>$ 0.13 $and$ Ax $<$ 0.13). 
In Figure \ref{fig:LxTxvsLopwithSubset}, 
it seems that clusters with very settled appearance 
tend to harbour the brighter BCGs compared to clusters very disturbed
appearance. 
Namely, the BCG luminosity correlation with cluster luminosity
or temperature seems to consists of two stronger parallel correlations for
disturbed and undisturbed clusters.
A K-S test along the BCG luminosity shows that the probability that the two
distributions are drawn from the same parent distribution is
1.14 $\times$ 10$^{-3}$.

\begin{figure}
 \resizebox{\hsize}{!}{\includegraphics[height=3cm,width=2cm,clip,angle=90]{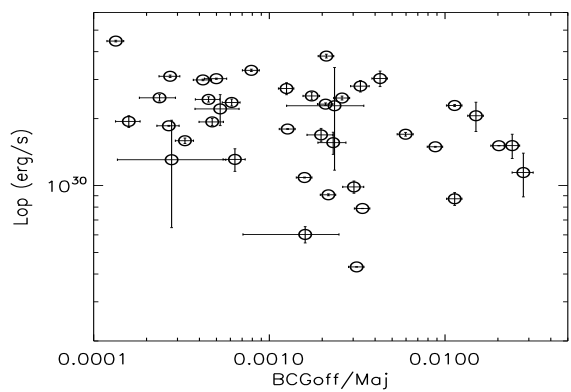}}
 \caption{
Luminosity of BCGs versus BCG offsets with respect to the cluster X-ray center.
}
\label{fig:OffsetvsLbcg}
\end{figure}

To further investigate this trend of dynamical influence in Figure \ref{fig:LxTxvsLopwithSubset}, we plotted
the luminosity of BCGs against the offset of the BCGs from the
X-ray center in Figure \ref{fig:OffsetvsLbcg}. 
The figure shows that BCGs at a larger offset from the X-ray center
exhibit a larger variation in their luminosity, and this variation
seems to be
always at `fainter' side  of the luminosity.
That is to say that
the figure shows a weak trend, with considerable scatter,
that the BCGs at smaller
offset from the X-ray center are brighter on the average 
(Spearman $\rho$ is -0.38, significance level = 1.4 $\times$ 10$^{-2}$).
The trends in both Figure \ref{fig:LxTxvsLopwithSubset} and \ref{fig:OffsetvsLbcg} 
may support the idea that the cluster-cluster merger 
may not be negligible compared to the scenario of the early collapse,
as the mechanism of BCG formation,
however, the more detailed discussion will be presented in
Sec. 6.

\color{black}

\section[]{Summary and Discussion}

We report an
investigation
of the relationship between the
clusters of galaxies and BCGs,
in particular, between the dynamical status of the clusters and the properties
of BCGs.

We find that:
1) BCGs show a wide variety of morphology in all of our objectively
determined morphological measures.
The variety is comparable to that of the bright red sequence galaxies, except in the
concentration  versus asymmetry plane. The bright red sequence galaxies show a much more compact
distribution  of morphology in that plane.
The second brightest, or the third brightest galaxies show a
similar morphological variety compared to the bright red sequence galaxies
in the concentration versus asymmetry plane, although the peak of distribution
seems slightly off to the lower concentration values
relative to that of the bright red sequence galaxies.
 The result seems to suggest that
 we have a continuous variation of morphology
 between BCG, BCG2, and  BCG3,
 rather than  a clear separation of morphological characteristics
 between the BCG and the rest of
 the bright galaxies.
2) The offset of the BCG position relative to the cluster center is correlated to
the possible cluster dynamical status, defined by cluster X-ray properties, in such a way that, inside dynamically
unsettled clusters, the BCGs tend to be more offset from the center of the
global cluster potential well.
3) The luminosity of the BCGs are weakly correlated to the cluster X-ray
luminosity or X-ray temperature, in a similar manner with cluster
scaling relation, but with considerable scatter. The scatter seems to be
much larger
than the previously reported relationship between the BCG luminosity
and X-ray luminosity or temperature,
when the cluster sample extends beyond nearby
X-ray bright clusters.
4) Effect of cluster dynamical status on the luminosity of the BCGs
seems to be comparable to
the effect related to the  cluster mass in such a way that the
dynamically stable clusters tend to harbour brighter BCGs than those of
unsettled clusters.

BCGs show morphology
distinctively different from that of the bright red sequence galaxies in the Cg-Ag plane. 
The origin of this difference may be due to the fact  that BCG galaxies tend to have
many superposed small galaxies that BCGs are perhaps in the process of accreting, rather than the nature of BCGs themselves that may be altered by 
the influence of local environments on BCGs 
\citep[e.g.][]{edwards2012close,lidman2013importance,burke2013growth}.
Namely, if the differences we find about the BCGs are indeed
due to the superposed small galaxies, then the difference between
the BCGs and other bright galaxies can be attributed, if not all, to the
`locational' difference rather than the fundamental difference in the nature
of the galaxies.
That is to say,
if we artificially superpose the image of, for example, the second or third brightest
galaxies near the cluster center by bringing them from outskirts to the center, at least some, if not all,
of the `BCG properties'  may 
be reproduced. 
This fact actually leads to  a very fundamental problem that can apply to 
any study of very extended objects: there may be no fundamentally trivial way to decouple an 
object itself 
and its immediate surroundings, and perhaps it is necessary to understand them in
one big picture of co-evolution of the object and the environments.

Our Figure \ref{fig:ExvsBCGoffset} and \ref{fig:CxvsBCGoffset} suggest that the BCG offset from the cluster center may
be correlated to the cluster X-ray morphology. This result implies
that the BCG offset from the cluster global potential well is sensitive
to the
cluster dynamical status, and that is consistent with previous studies
investigating BCG offset from the optically determined cluster center 
among nearby clusters 
\citep[e.g.][]{ostriker1975another,merritt1985distribution,katayama2003properties}.
However, just as other previous studies,
the measurement of BCG offset may not be completely independent
from the various measures characterizing 
the dynamical status of the clusters, i.e. two measurements are not
completely orthogonal parameters,
and therefore the physics behind the correlation between the BCG offset
and the dynamical status of clusters should be interpreted with caution.
Nevertheless, the fact remains that BCG offset can be a good measure to
characterize the possible cluster dynamical status.

The fact that there is a weak correlation between the BCG luminosity
and the X-ray luminosity or X-ray temperature of host cluster
is consistent with previous studies reporting similar weak correlation
between the X-ray cluster properties and BCG luminosity 
\citep[e.g.][]{schombert1987structure,edge1991exosat,brough2002evolution,katayama2003properties,stott2012xmm}.
Our result is also in agreement with more traditional optical investigation
where the BCG luminosity was compared to the optical properties of clusters
\citep[e.g.][]{oemler1976structure,schombert1987structure,lin2004k}.
The strengths of the correlations
varied from one study to another, depending on the cluster measures and cluster 
samples of the study. 
Unfortunately, most of previous studies were predominantly
using the low redshift bright cluster samples, that preferentially contained dynamically settled clusters.
Therefore, the effect of cluster dynamical status on the luminosity (or other
properties) of BCGs were hardly addressed. 

Our Figure \ref{fig:LxTxvsLopwithSubset} 
shows that undisturbed looking clusters statistically
harbour bright BCGs compared to those of disturbed clusters,
implying that the effect of the dynamical status of clusters on the BCG
luminosity may not be negligible.
The result is qualitatively consistent with the previous studies
revealing correlations between the `luminosity gap' and cluster 
structure, where they reveal the trend that dynamically settled
clusters tend to harbour dominant BCG \citep[e.g.][]{smith2010locuss,chon2012statistics}.  
 
Note that very high Cx clusters can contain `cool core' clusters   
\citep[e.g.][]{peterson2006x}.  That is because of indirect selection effect 
that
dynamically settled clusters tend to harbour the cool core, 
as well as the direct effect that Cx measure itself may be
sensitive to the small scale X-ray profile  
associated with the cool core clusters near the center.
This is the reason behind why the cluster morphology are characterized
by both Cx and Ax in Figure \ref{fig:LxTxvsLopwithSubset} 
to minimize, at least, the direct influence from 
cool core using Cx alone.

Figure \ref{fig:OffsetvsLbcg} is qualitatively 
consistent with Figure \ref{fig:LxTxvsLopwithSubset}, 
in such a way that  
dynamically settled clusters tend to harbour brighter BCGs.
\citet{stott2012xmm} reported a similar weak correlation 
between BCG optical luminosity and BCG offset, but with considerable scatter. 
This larger scatter is likely to be originated from the large uncertainty
in determining their cluster centroids using shallow serendipitous observations with XMM-Newton.

What do these results in Figure \ref{fig:LxTxvsLopwithSubset} and \ref{fig:OffsetvsLbcg} 
imply about the formation and evolution of BCGs ?
The popular scenarios of BCG formation and evolution are ``galactic cannibalism"
\citep[e.g.][]{ostriker1975another,malumuth1984evolution}, 
and ``early collapse'' \citep[e.g.][]{merritt1984relaxation}. 
The fact that our BCG luminosity is correlated with cluster dynamical status
may imply, at least, that the ``early collapse'' is not the only dominant mechanism to control the BCG formation and evolution.

However, it is pointed out that a simple galaxy-galaxy cannibalism scenario may require
unusually short dynamical friction time scale to account for the observed
large luminosity of BCGs \citep[e.g.][]{merritt1984relaxation}. 
Indeed, \citet{katayama2003properties}
showed that, if their ``virial density'' could be interpreted as an age 
indicator
of a cluster, BCG luminosity might not be correlated well to the age of the cluster,
suggesting the possibility of being inconsistent with the simple galaxy-galaxy cannibalism scenario. 
Our dynamical indicators defined by X-ray morphology of clusters
are expected to be 
more sensitive
to a larger scale cannibalism, such as cluster-cluster merging, and
may 
cover a longer evolutionary time scale than 
the ``quiescent'' age indicator, such as the virial density.
We, therefore, suspect 
the fact that our BCG luminosity is correlated with our dynamical indicators
does not support the simple galaxy-galaxy cannibalism
model, but is more consistent with a larger/longer scale cannibalism as the mechanism of BCG formation and evolution, such as, cluster-cluster,
or group-group merging, that can take place throughout the entire 
course of the cluster evolution
even after the initial collapse \citep[e.g.][]{lin2004k}. 

Note however that, 
after the initial collapse of the BCGs, both cluster-cluster, and galaxy-galaxy
cannibalisms can take place.
To determine
the relative importance of three different mechanisms: early collapse,
galaxy-galaxy cannibalism, and cluster-cluster merger,
the investigation of the occurrence of distant
``fossil groups'' \citep[e.g.][]{jones2003nature}, or OLEG (X-ray overluminous elliptical galaxies) \citep[][]{vikhlinin1999x}, 
as well as numerical simulation exploiting a wider parameter range of
various
merging scales, is much needed.

\section*{Acknowledgments}
 This work is based in part on data collected at Subaru Telescope and obtained from the SMOKA, which is operated by the Astronomy Data Center, National Astronomical Observatory of Japan. This research has made use of data obtained from the Chandra Data Archive, and software provided by the Chandra X-ray Center (CXC) in the application packages CIAO. 
 We acknowledge the referee's comments, which improved the manuscript.
 The work is partially supported by the National Science Council of Taiwan under the grant NSC 99-2112-M-003-001-MY2.

\bibliographystyle{mn2e}
\bibliography{reflist}

\end{document}